\documentclass[journal,10pt]{IEEEtran}
\usepackage{graphicx, epstopdf}
\usepackage{amssymb}
\usepackage{amsmath}
\usepackage{amsthm}
\usepackage{mathrsfs}
\usepackage{cite}
\usepackage{color}
\usepackage[acronym]{glossaries}
\usepackage[T1]{fontenc}
\usepackage[latin9]{inputenc}
\usepackage{array}
\usepackage{textcomp}
\usepackage{multirow}
\usepackage{soul}

\newacronym{henet}{HetNet}{heterogeneous network}
\newacronym{tr}{TR}{time reversal}
\newacronym{isi}{ISI}{inter-symbol interference}
\newacronym{bs}{BS}{base station}
\newacronym{mbs}{MBS}{macrocell base station}
\newacronym{fbs}{FBS}{femtocell base station}
\newacronym{fbss}{FBS(s)}{femtocell base stations}
\newacronym{mu}{MU}{macrocell user}
\newacronym{fu}{FU}{femtocell user}
\newacronym{cir}{CIR}{channel impulse response}

\newacronym{cr}{CR}{cognitive user}
\newacronym{cr-tx}{CR-Tx}{cognitive transmitter}
\newacronym{cr-rx}{CR-Rx}{cognitive receiver}
\newacronym{fcc}{FCC}{Federal Communications Commission}
\newacronym{dsa}{DSA}{Dynamic Spectrum Access}
\newacronym{ssa}{SSA}{Static Spectrum Access}
\newacronym{rf}{RF}{radio frequency}
\newacronym{csd}{CSD}{cyclic spectrum density}
\newacronym{sbs}{SBS}{secondary base station}
\newacronym{pbs}{PBS}{primary base station}
\newacronym{ack}{ACK}{acknowledgement}
\newacronym{arq}{ARQ}{automatic repeat request}
\newacronym{fifo}{FIFO}{first in first out}
\newacronym{iid}{i.i.d}{identically independent distributed}
\newacronym{inid}{i.n.i.d}{independent but not necessarily identically distributed}
\newacronym{nack}{NACK}{Not-acknowledgement}
\newacronym{rtt}{RTT}{round-trip-time}
\newacronym{pip}{PIP}{peak interference power to noise ratio}
\newacronym{crn}{CRN}{cognitive radio network}
\newacronym{ccrn}{CCRN}{cognitive cooperative radio network}
\newacronym{su}{SU}{secondary user}
\newacronym{sr}{SR}{secondary relay}
\newacronym{su-tx}{S-Tx}{secondary transmitter}
\newacronym{su-rx}{S-Rx}{secondary receiver}
\newacronym{pu}{PU}{primary user}
\newacronym{pu-tx}{P-Tx}{primary transmitter}
\newacronym{pu-rx}{P-Rx}{primary receiver}
\newacronym{af}{AF}{amplify and forward}
\newacronym{df}{DF}{decode-and-forward}
\newacronym{cdf}{CDF}{cumulative distribution function}
\newacronym{pdf}{PDF}{probability density function}
\newacronym{sinr}{SINR}{signal-to-interference-plus-noise ratio}
\newacronym{snr}{SNR}{signal-to-noise ratio}
\newacronym{awgn}{AWGN}{additive white Gaussian noise}
\newacronym{csi}{CSI}{channel state information}
\newacronym{qos}{QoS}{quality of service}
\newacronym{rv}{RV}{random variable}
\newacronym{simo}{SIMO}{single-input multiple-output}
\newacronym{mimo}{MIMO}{multiple-input multiple-output}
\newacronym{miso}{MISO}{multiple-input single-out}
\newacronym{sc}{SC}{selection combining}
\newacronym{tas}{TAS}{transmit antenna selection}
\newacronym{pinr}{PIP}{peak-interference-power-to-noise ratio}
\newacronym{cren}{CReN}{cognitive relay network}
\newacronym{d2d}{D2D}{device-to-device communication}
\newacronym{eav}{EAV}{eaversdropper}
\newacronym{sep}{SEP}{symbol error probability}
\newacronym{lte-a}{LTE-A}{long-term evolution-advanced}

\newcommand{\xv}{\mathbf{x}}
\newcommand{\uv}{\mathbf{u}}
\newcommand{\gv}{\mathbf{g}}
\newcommand{\pv}{\mathbf{p}}

\title{Downlink Power Optimization for Heterogeneous Networks with Time Reversal-based
Transmission under Backhaul Limitation}
\author{
        {
        Ha-Vu Tran$^{1}$, Georges Kaddoum$^{1}$, Hung Tran$^{2}$, and Een-Kee Hong$^{3}$
        }

\thanks{
$^1$Ha-Vu Tran and Georges Kaddoum are with University of
Qu\'{e}bec, \'{E}TS engineering school, LACIME Laboratory, 1100 Notre-Dame west, H3C 1K3, Montreal, Canada.
Email: \{ha-vu.tran.1@ens.etsmtl.ca, georges.kaddoum@etsmtl.ca.\}

$^2$Hung Tran is with School of Innovation, Design and Engineering, M\"{a}lardalen University, 721 23  V\"{a}ster{\aa}s, Sweden. Email: \{tran.hung@mdh.se\}

$^3$Een-Kee Hong is with School of Electronics and Information, Kyung
Hee University,  Yongin 449-701, South Korea.
Email: \{ekhong@khu.ac.kr\}

This work has been supported by NSERC discovery grant 435243 - 2013.
}
 }

\begin{document}
    \maketitle
\begin{abstract}
In this paper, we investigate an application of two different beamforming techniques and propose a novel downlink power minimization scheme for a two-tier heterogeneous network (HetNet) model.
In this context, we employ time reversal (TR) technique to a femtocell base station (FBS) whereas we assume that a macrocell base station (MBS) uses a zero-forcing-based algorithm and the communication channels are subject to frequency selective fading. Additionally, HetNet's backhaul connection is unable to support a sufficient throughput for signaling information exchange between two tiers. Given the considered HetNet model, a downlink power minimization scheme is proposed, and closed-form expressions concerning the optimal solution are provided, taking this constraint into account. Furthermore, considering imperfect channel estimation at TR-employed femtocell, a worst-case robust power minimization problem is formulated. By devising TR worst-case analysis, this robust problem is transformed into an equivalent formulation that is tractable to solve. The results presented in our paper show that the TR technique outperforms the zero-forcing one in the perspective of beamforming methods for femtocell working environments. Finally, we validate the proposed power loading strategy for both cases of perfect and imperfect channel estimations.

\end{abstract}

\section{Introduction}\label{sec:1-Introduction}
Recently, \gls{henet} has been considered as a promising solution to enhance the throughput and to overcome the drawbacks of traditional cellular networks \cite{Zah2013,Ekram2015,StefanoBuzzi2016}, such as the inefficient usage of spectrum and dynamic spectrum access. According to the \gls{henet} concept, the macrocell serves a large number of users in a wide area  while low power cells such as femtocells, picocells and microcells  handle a smaller number of users. Following this approach, not only the coverage range is expanded but also the throughput and reliability can be improved significantly.
More specifically, the works reported in \cite{Zah2013, Ekram2015,StefanoBuzzi2016,Andrews2012} have investigated a HetNet model in which an original cellular network is decomposed into multi-tier
networks, and each tier is responsible for a specific zone. These approaches
 have expanded the coverage ranges over the dead zones and hot zones
of traditional cellular networks. Therefore, the \emph{femtocell} is considered as one of the most cost-efficient provisioning for cellular network services \cite{Ben2011}.

Regarding the radio environment, the signal power is often degraded due to path-loss effects and multipath propagation, and such an issue becomes more severe in the frequency selective fading.
On the other hand, many techniques have been employed to mitigate the adverse effects of frequency selective channels such as: equalizers, \gls{mimo}  and  \gls{tr} techniques. When applied to wireless receivers, the first two techniques provide significant enhancement to the received signal-to-noise ratio (SNR). However, from the implementation point of view, these techniques are high-cost and require complex equipment which make them less interesting to be used at the subscriber end where limited energy and processing resources are the major constraints.

However, a special class of beamforming technique, namely the \gls{tr}, which was mainly used in acoustics and underwater communication systems, has been proposed in \cite{Tran-Ha2015,Kai2010,Bou2013,Maaz2015,YanChen2016, Pit2014,Siaud2015,Carlos2015} to wireless communications, e.g. ultra-wideband, large-scale antenna, and millimeter-wave systems. 
This techique provides a promising solution to save the processing cost and to combat the adverse effects of frequency selective fading channels.
Benefiting from the reciprocal properties of wireless channels, the TR technique principles rely on using the time-reversed form of a \gls{cir} to pre-filter the transmitted signal which leads to the power convergence of this latter in the time and space domains at the receiver side. Specifically, the \gls{cir} at the transmitter side is estimated by virtue of a pilot signal sent from the receiver.

Particularly, some works have addressed the designs of TR beamforming. 
In \cite{Bou2013}, the authors have provided an analysis of the TR technique for green radio communications employed to
WiFi-certified technologies. The work \cite{Carlos2016} proposes three forms of space-time block diagonalization on the platform of the TR technique.
Further, Yang {\it et.al.} \cite{Yan2013} propose a novel TR waveform to maximize the sum-rate of a multi-user system. 
In addition, the paper \cite{Eunchul2015} introduces a design of a TR-based waveform using predistortion to combat \gls{isi}.
In fact, encouraging results obtained in \cite{YanChen2016,Bou2013,Maaz2015} show that the TR-based transmission is an ideal paradigm for green wireless communications.
Moreover, the experiments in \cite{Hen2004, Ngu2006} confirm that the \gls{tr} technique is feasible for broadband systems including femtocell networks.

In our paper, we focus on studying a realistic scenario of \gls{henet} system consisting of \gls{mbs} and \gls{fbs} and their users under backhaul limitation. The different channels in this network are subject to frequency selective fading. In our model, it is assumed that different cellular stations are equipped with multiple antennas whereas each receiver has one antenna due to the limited resources at the user end. 
Conventionally, a central controller, likely \gls{mbs}, is responsible to compute the beamformers and power vectors for each \gls{bs} located in the \gls{henet}. Hence, this process requires a solid backhaul connections that must be always available to accommodate the central controller with all \gls{csi} of the different users located in different cells \cite{Andrews2012, Tolli2011}. However, in a realistic case that backhaul connection {endures congestion} in which obtaining sufficient amount of \gls{csi} might become infeasible. 
Therefore, given the system model, this work aims at seeking solutions for the question that {\it how to mitigate the frequency selectivity of fading channels, and to deal with the limited backhaul connection while taking the processing burden of the macrocell and the transmit power restriction of the femtocell network into account?} In the following, the potential proposed methods are discussed.

In this vein, one of our novelties consists of applying zero-forcing and \gls{tr} techniques to \gls{mbs} and \gls{fbs} respectively to combat channel selectivity and to enhance network performance. 
In fact, zero-forcing is one of the most efficient beamforming techniques, and it is an interesting solution for macrocell networks \cite{Roze2015,Vutobepublished}.
However, in a femtocell working environment where FBS's transmit power is limited, zero-forcing might not be a promising approach due to the transmit power restriction and hardware limitation \cite{Roze2015,Peel2005, Zah2013}. In this case, the \gls{tr} which offers an alternative low-cost beamforming technique, is proposed to provide a better system performance for femtocell networks. 

Moreover, 
we propose a novel optimal power allocation method, assuming that the backhaul connection may only convey a limited throughput for signaling exchange. 
In single-tier multi-cell networks, the concept of cross-interference management has been introduced to deal with backhaul limitation \cite{Tolli2011}.
However, this approach might not be applied to multi-tier HetNets directly since \glspl{mu} and \glspl{fu} have different priorities. 
Besides, there are several previous works addressing the issue of backhaul limitation for HetNets \cite{Joe2011, Duy2015}. In principle, these works focus on splitting the conventional optimization problem into two subproblems (i.e. one for the macrocell and the other for the femtocell) in which solutions can be achieved with a reduced amount of required CSI. In our work, decoupling the original problem is adopted in a different manner to deal with backhaul limitation. In particular, the proposed scheme only requires the minimized cross-tier interference sent from the femtocell. This latter reduces the signaling overhead in the network compared with the scheme proposed in \cite{Joe2011}, and releasing the FUs from the task of measuring cross-interference caused by the MBS compared with the another work \cite{ Duy2015}. Moreover, importantly, our scheme can control the priority of MUs by using a preset threshold of the cross-tier interference that the MBS causes to the FUs.
On this basis, the network operator can flexibly manage the overall network performance.
Especially, we solve the considered macrocell and femtocell problems by devising optimal closed-form solutions which do not appear in the literature.

{
Furthermore, most of previous publications \cite{Bou2013,Maaz2015, Eunchul2015, Carlos2016} address the designs of TR beamforming under the assumption of perfect channel estimations. There are no existing works which consider the worst-case robust beamforming for the TR technique.}
Given this conern, the robust design is formulated into a non-convex problem.
To transform such a problem into a tractable formulation to solve, the effects of channel estimation errors (CEE) on TR-based systems are analyzed in terms of the worst-case boundaries of desired signal and interference components.
Especially, to tackle this case, a well-known Young's inequality \cite{You1912} is able to bring an efficient solution dealing with the boundaries of \gls{isi} and co-tier interference. However, a novel tighter boundary formulation is derived to enhance the performance of power allocation strategy. On this basis, the robust optimization problem is relaxed into a convex problem that can be solved by closed-form expressions.

According to the discussed content, the main contributions of this work are summarized as follows
\begin{itemize}
\item The application of \gls{tr} technique for the femtocell network is proposed.
\item A novel downlink optimization method dealing with limited backhaul connections is provided.

\item Closed-form optimal solutions are derived for both the downlink power minimization problems of macrocell and femtocell.

\item A robust worst-case power allocation problem of TR-employed femtocell under the effects of imperfect \gls{csi} is analyzed. 
\end{itemize}

Given this outline, the remainder of this paper is organized as follows: 
In Section \ref{sec:SystemModel}, the system model is described. 
The beamforming designs for MBS and FBS are discussed in Section \ref{sec:Beamformingdesigns}.
In Section \ref{sec:DistributedPowerAllocation}, the proposed power allocation
manner is presented. In Section \ref{sec:WorstCase}, the worst-case robust problem is formulated and analyzed. In Section \ref{sec:NumericalResult}, numerical results and discussions are provided. Finally, concluding remarks are put forward in Section \ref{sec:Conclusion}.

\textit{Notation:} The notation $\mathbb{R}_{+}^{m}$
and $\mathbb{C}^{m\times n}$ denote the sets of $m$-dimensional nonnegative real vector
and $m \times n$ complex matrix, respectively. The boldface
lowercase $\textbf{a}$ and uppercase $\textbf{A}$ indicate vectors and matrices, respectively.
The superscripts ${\bf A}^{T}$ and ${\bf A}^{H}$ represent the transpose
and transpose conjugate, respectively. In addition, symbols $\left|.\right|$, $\left\Vert .\right\Vert $,
and $\left\Vert .\right\Vert _{1}$ stand for the absolute value,
vector Euclidean norm and vector $l_{1}$-norm, respectively. For
a complex value, we denote $\Re\{.\}$ and $\Im\{.\}$ to be the real and imaginary part, respectively.

\section{System Model}\label{sec:SystemModel}
\begin{figure}[!]
\fbox{\includegraphics[width=0.48\textwidth]{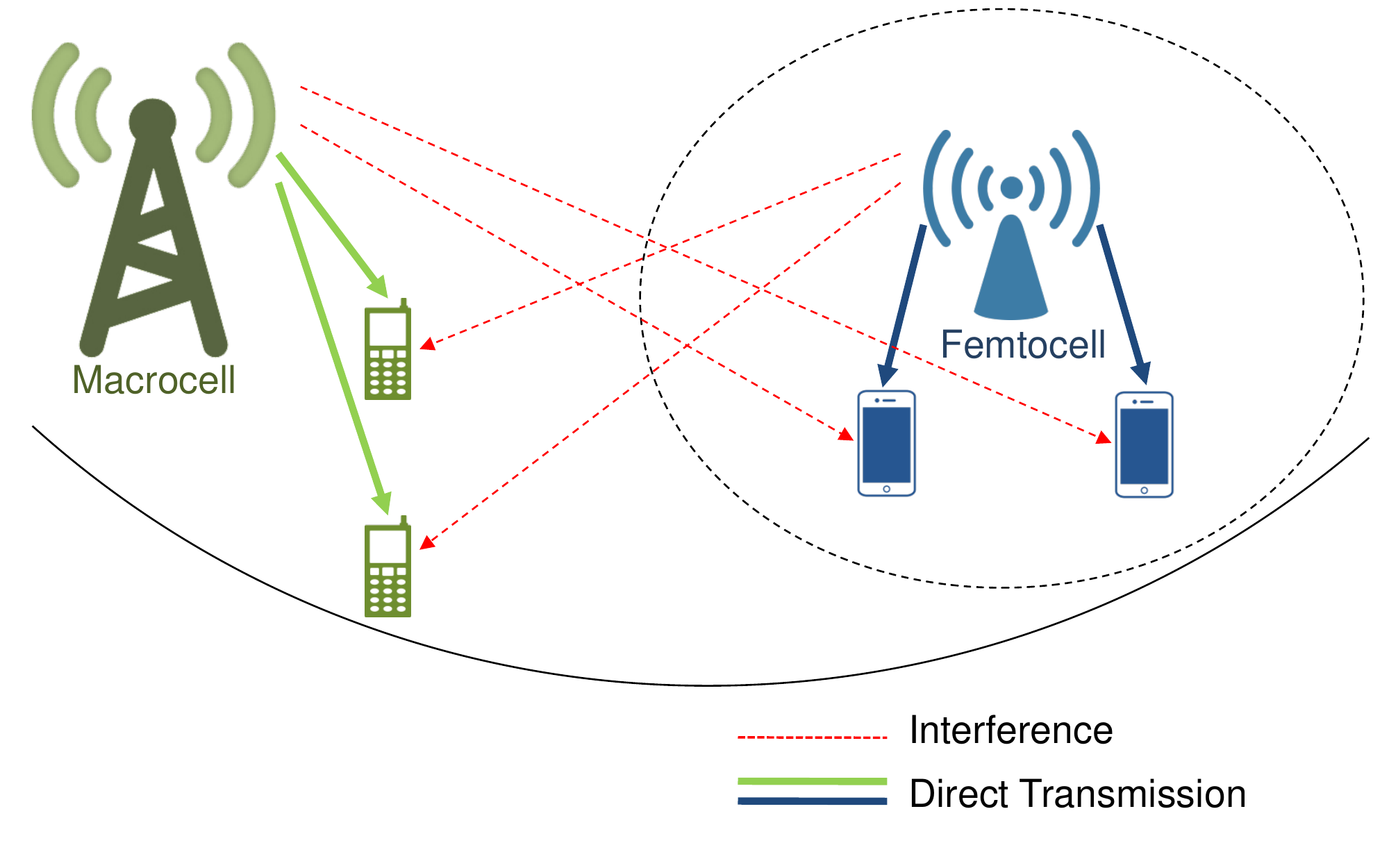}}
   \caption{
       A two-tier system model including a macrocell and a femtocell.
    }
    \label{fig:SystemModel}
\end{figure}

As shown in Fig.~\ref{fig:SystemModel}, a two-tier \gls{henet} system including one \gls{mbs} and one \gls{fbs} is considered.
For convenient notation, we denote the MBS as $\mathcal{B}_0$ and the FBS as $\mathcal{B}_1$.
We assume that $\mathcal{B}_k$ ($k=\{0,1\}$) is equipped with $M_{k}$ antennas and serves $N_k$ users.
On the other hand, FUs and MUs are equipped with a single antenna and a single tap diversity combiner.

Let ${\bf h}_{ij}^{kr}\in\mathbb{C}^{L\times1}$
($k,\; r=\{0,1\}$, \textbf{ $0\le i\le M_{k},~ 0\le j\le N_{r}$) } denote
the \gls{cir} between the $i^\text{th}$ transmission antenna of {$\mathcal{B}_k$} and
the $j^\text{th}$ user of {$\mathcal{B}_r$}. Moreover, $L$ denotes the
maximum length of each \gls{cir},  and the superscript $k$ is used to represent superscript
$kk$ for convenience in notation, (e.g. ${\bf h}_{j}^{0}$ is used to denote ${\bf h}_{j}^{00}$).

Therefore, the transmitted signals at {$\mathcal{B}_0$} and {$\mathcal{B}_1$} can be formulated, respectively, as 
\begin{align}
\label{eq:x_n0}
   \xv_{n}^{0}&=\sqrt{p_{n}^{0}}
          \left[
            \begin{array}{ccc}
                  \uv_{{1}n} & \ldots &  \uv_{M_{0}n}
            \end{array}
           \right]^{}s_{n}^{0}, {(\xv_{n}^{0} \in\mathbb{C}^{L \times M_0} )}
   \\
   \label{eq:x_j1}
   \xv_{j}^{1}&=\sqrt{p_{j}^{1}}
            \left[
            \begin{array}{ccc}
                 \gv_{{1}j} & \ldots & \gv_{M_{1}j}
            \end{array}
            \right]^{}s_{j}^{1}, {(\xv_{j}^{1} \in\mathbb{C}^{L \times M_1} )}
\end{align}
where $s_{n}^{0}$ and $s_{j}^{1}$ {are scalars representing} the unit power transmitted
symbols for the $n^\text{th}$ \gls{mu} and the $j^\text{th}$ \gls{fu}, respectively.
We define ${\pv}^{k}=\left[\begin{array}{cccc}
{p_{1}^{k}} & {p_{2}^{k}} & \ldots & {p_{N_{k}}^{k}}\end{array}\right]^{T}\in\mathbb{R}_{+}^{N_{k}\times1}$ as the transmit power vector of the $\mathcal{B}_k$. 
Furthermore, ${\bf u}_{mn}\in\mathbb{C}^{L\times1}$ is the
beamformer for the $n^\text{th}$ \gls{mu} used at the $m^\text{th}$
transmission antenna, and $\gv_{ij} \in \mathbb{C}^{L\times1}$ is
the beamformer for the $j^\text{th}$ \gls{fu} employed at the $i^\text{th}$ transmit
antenna.
Specifically, ${\bf u}_{mn}$ follows a zero-forcing-based algorithm whereas $\gv_{ij}$ has the formulation of \gls{tr} beamformer. The design of two such beamforming vectors is thoroughly discussed in Section IV. A and Section IV. B.

In our paper, we consider the downlink communication scenario in which the \gls{mbs} and the \gls{fbs} transmit their signals to their corresponding users simultaneously and none of these communicate with its \gls{bs} during this phase of communication. Accordingly, the received signal at the $n^\text{th}$ MU can be written as
\begin{align}
{\bf y}_{n}^{0} &= \sum\limits _{m=1}^{M_{0}}\sqrt{p_{n}^{0}}{\bf u}_{mn}*{\bf h}_{mn}^{0}s_{n}^{0}+ \sum\limits _{{n'=1\atop n'\ne n}}^{N_{0}}\sum\limits _{m=1}^{M_{0}}\sqrt{p_{n'}^{0}}{\bf u}_{mn'}*{\bf h}_{mn}^{0}s_{n'}^{0}
\nonumber \\
\label{eq:y_n}
 &+\sum\limits _{j=1}^{N_{1}}\sum\limits _{i=1}^{M_{1}}\sqrt{p_{j}^{1}}{\bf g}_{ij}*{\bf h}_{in}^{10}s_{j}^{1} +{\bf n}_{M}, {({\bf y}_{n}^{0} \in\mathbb{C}^{(2L-1) \times 1} )}
\end{align}
where ${\bf n}_{M}$ is \gls{awgn}, and $*$ is the  convolution operator. It is noted that the first term in \eqref{eq:y_n} is the received
signal of the $n^\text{th}$ MU while the second term is the co-tier
interference in the macrocell and the third term is the cross-tier interference {from the femtocell}.

Hence, {at the $n^\text{th}$ MU}, we define $ P_{(sig)}{}_{n}^{0}$, ${P_{(isi)}{}_{n}^{0}}$, ${P_{(co)}{}_{n}^{0}}$, and ${P_{(cross)}{}_{n}^{0}}$ as the power of desired
signal, \gls{isi}, co-tier interference power and cross-tier interference from femtocell, respectively, as indicated in \eqref{eq:P_sig_n0}-\eqref{eq:P_cross_n0}.
\begin{align}
    \label{eq:P_sig_n0}
        P_{(sig)}{}_{n}^{0}&=
        \left|\left(\sum\limits _{m=1}^{M_{0}}\sqrt{p_{n}^{0}} {{\bf u}_{mn}*{\bf h}_{mn}^{0}}\right)\left[\alpha\right]\right|^{2},
\end{align}
\begin{align}
        \label{eq:P_isi_n0}
        {P_{(isi)}{}_{n}^{0}}&=\sum\limits _{l\ne\alpha}^{2L-1}\left|\left(\sum\limits _{m=1}^{M_{0}}\sqrt{p_{n}^{0}}{{\bf u}_{mn}*{\bf h}_{mn}^{0}}\right)\left[l\right]\right|^{2},
\end{align}
\begin{align}
        \label{eq:P_co_n0}
        {P_{(co)}{}_{n}^{0}}&=\sum\limits _{{n'=1\atop n'\ne n}}^{N_{0}}\left\Vert \sum\limits _{m=1}^{M_{0}}\sqrt{p_{n'}^{0}}{\bf u}_{mn'}*{\bf h}_{mn}^{0}\right\Vert ^{2},
 \end{align}
\begin{align}
        \label{eq:P_cross_n0}
        {P_{(cross)}{}_{n}^{0}}&=\sum\limits _{j=1}^{N_{1}}\left\Vert \sum\limits _{i=1}^{M_{1}}\sqrt{p_{j}^{1}}{\bf g}_{ij}*{\bf h}_{in}^{10}\right\Vert ^{2},
\end{align}
where $\alpha$ represents the position of the selected tap.

Accordingly, the \gls{sinr} at the $n^\text{th}$
\gls{mu}  can be formulated as 
\begin{align}
\label{eq:SINR_n0}
    &\text{SINR}_{n}^{0}\left(\alpha,{\bf p}^{0},\left\{ {\bf u}_{mn}\right\} _{n=1}^{N_{0}},P_{(cross)}{}_{n}^{0}\right)\nonumber\\
    &=
    \frac{{P_{(sig)}{}_{n}^{0}}}
    {{P_{(isi)}{}_{n}^{0}}+{P_{(co)}{}_{n}^{0}}+P_{(cross)}{}_{n}^{0}+\left\Vert {\bf n}_{M}[\alpha]\right\Vert ^{2}},
\end{align}

On the other hand, the received signal at the $j^\text{th}$ \gls{fu} can be expressed as
\begin{align}
{\bf y}_{j}^{1} & =  \sum\limits _{i=1}^{M_{1}}\sqrt{p_{j}^{1}}{\bf g}_{ij}*{\bf h}_{ij}^{1}s_{j}^{1}+\sum\limits _{{j'=1\atop j'\ne j}}^{N_{1}}\sum\limits _{i=1}^{M_{1}}\sqrt{p_{j'}^{1}}{\bf g}_{ij'}*{\bf h}_{ij}^{1}s_{j'}^{1}\nonumber
\\
\label{eq:y_j1}
 & +\sum\limits _{n=1}^{N_{0}}\sum\limits _{m=1}^{M_{0}}\sqrt{p_{n}^{0}}{\bf u}_{mn}*{\bf h}_{mj}^{01}s_{n}^{0}+{\bf n}_{F}, {({\bf y}_{j}^{1} \in\mathbb{C}^{(2L-1) \times 1} )}
\end{align}
where ${\bf n}_{F}$ is \gls{awgn}. Here, the first term in \eqref{eq:y_j1} is the received
signal for the $j^\text{th}$ \gls{fu}, the second term is the co-tier interference
in the femtocell, and the third term is the cross-tier interference from
the macrocell.

 Similarly, $ P_{(sig)}{}_{j}^{1}$, ${P_{(isi)}{}_{j}^{1}}$, ${P_{(co)}{}_{j}^{1}}$, and ${P_{(cross)}{}_{j}^{1}}$, i.e. \eqref{eq:P_sig_j1}-\eqref{eq:P_cross_j1}, represent the power of the desired
signal, \gls{isi}, co-tier interference and cross-tier interference from macrocell, respectively. 

 \begin{align}
 \label{eq:P_sig_j1}
 P_{(sig)}{}_{j}^{1}
 &=
 \left|\left(\sum\limits _{i=1}^{M_{1}}\sqrt{p_{j}^{1}}{\bf g}_{ij}*{\bf h}_{ij}^1\right)\left[\beta\right]\right|^{2},
 \end{align}
\begin{align}
  \label{eq:P_isi_j1}
 {P_{(isi)}{}_{j}^{1}}&=\sum\limits _{{l=1\atop l\ne\beta}}^{2L-1}\left|\left(\sum\limits _{i=1}^{M_{1}}\sqrt{p_{j}^{1}}{\bf g}_{ij}*{\bf h}_{ij}^{1}\right)\left[l\right]\right|^{2},
 \end{align}
\begin{align}
 \label{eq:P_co_j1}
 {P_{(co)}{}_{j}^{1}}&=\sum\limits _{{j'=1\atop j'\ne j}}^{N_{1}}\left\Vert \sum\limits _{i=1}^{M_{1}}\sqrt{p_{j'}^{1}}{\bf g}_{ij'}*{\bf h}_{ij}^{1}\right\Vert ^{2},
 \end{align}
\begin{align}
 \label{eq:P_cross_j1}
 {P_{(cross)}{}_{j}^{1}}&=\sum\limits _{n=1}^{N_{0}}\left\Vert \sum\limits _{m=1}^{M_{0}}\sqrt{p_{n}^{0}}{\bf u}_{mn}*{\bf h}_{mj}^{01}\right\Vert ^{2},
 \end{align}
where $\beta$ denotes the position of selected tap. Note that the problem of selecting the values of $\alpha$ and $\beta$ is discussed in the following Section IV.

Thus, the \gls{sinr} of the $j^\text{th}$ \gls{fu} can be expressed as 
\begin{align}
\label{eq:SINR_j1}
&\text{SINR}_{j}^{1}\left(\beta,{\bf \; p}^{1},\;\left\{ {\bf g}_{ij}^{1}\right\} _{i=1,j=1}^{M_{1},N_{1}}, {P_{(cross)}{}_{j}^{1}}\right) \nonumber\\
&=\frac{{P_{(sig)}{}_{j}^{1}}}{{P_{(isi)}{}_{j}^{1}}+{P_{(co)}{}_{j}^{1}}+{P_{(cross)}{}_{j}^{1}}+\left\Vert {\bf n}_{F}[\beta]\right\Vert ^{2}}.
\end{align}

\section{Beamforming Designs for the MBS and the FBS}\label{sec:Beamformingdesigns}
This section provides the beamforming designs for FBS and MBS over frequency selective fading channels. In more details, MBS uses a zero-forcing-based algorithm whereas FBS employs the TR technique.

\subsection{Beamformer design following the zero-forcing technique for the MBS}
{It is well-known that the zero-forcing technique mainly aims to suppress the interference components. Also, such a technique leads to the fact that the desired signal strength might only reach a limited level. 
However, this issue can be overcome by using a high transmit power. Given this concern, the signal strength can be significantly improved whereas the interference is cancelled.}
Due to the avaibility of power at MBS, zero-forcing beamforming technique is preferred as an efficient solution in such an environment.
On the other hand, user receivers take only one sample at a particular tap.
For the $l^\text{th}$ case, we specifically treat the $l^\text{th}$ tap as the desired tap whereas the other taps can be considered as the \gls{isi} taps respectively. The corresponding beamformer must be designed following the zero-forcing scheme to suppress both ISI and co-tier interference. Thereupon, we obtain $(2L-1)$ relevant beamformers. Finally, we finger out the beamformer among these candidates that yields the best SINR at the user. In the following content, the zero-forcing-based algorithm is discussed in details.

Let us start with a beamforming design based on the well-known zero-forcing technique over macrocell environments by re-arranging values of $\left\{{\bf h}_{mn}^{0}[l]\right\}_{m=1}^{M_0}$ into a new form ${\bf \bar{h}}_{ln} \in \mathbb{C}^{1\times M_{0}}$ as follows
\begin{align}
{\bf \bar{h}}_{ln}=\left[\begin{array}{cccc}
{h}_{1n}^{0}\left[l\right] & {h}_{2n}^{0}\left[l\right] & \ldots & {\rm h}_{M_{0}n}^{0}\left[l\right]\end{array}\right].
\end{align}
Further, we define a matrix ${\bf \bar{H}}_{n}\in\mathbb{C}^{\left(2L-1\right)\times M_{0}L}$
as
\begin{align}
\begin{aligned}{\bf \bar{H}}_{n}= & \left[\begin{array}{cccc}
{\bf \bar{h}}_{1n} & \bf 0 &  & \bf 0\\
{\bf \bar{h}}_{2n} & {\bf \bar{h}}_{1n}\\
\vdots & {\bf \bar{h}}_{2n}\\
{\bf \bar{h}}_{Ln} & \vdots & \ddots & {\bf \bar{h}}_{1n}\\
 & {\bf \bar{h}}_{Ln} &  & {\bf \bar{h}}_{2n}\\
 &  & \ddots & \vdots\\
\bf 0 &  &  & {\bf \bar{h}}_{Ln}
\end{array}\right].\end{aligned}
\end{align}
{Note that the matrix ${\bf \bar{H}}_{n}$ is derived following the formulation of the Sylvester matrix of $({\bf h}_{mn}^{0})^T$ [pp. 28, \cite{Trott2007}]}. For each sampled tap $\bar{\alpha}^\text{th}$, let us define ${\bar{\bf u}}_{mn,\bar{\alpha}} \in \mathbb C^{L\times 1}$ as the $\bar{\alpha}^\text{th}$ candidate for the beamformer ${{\bf u}}_{mn}$.

On the basis of zero-forcing principle, the beamformer of MBS can be derived according to a computation as follows
\begin{align}\label{eq:VectorW1}
{ \text{vec}(\left[{\bar{\bf u}}_{1n,\bar{\alpha}} \quad {\bar{\bf u}}_{2n,\bar{\alpha}} \quad  \ldots \quad { \bar{\bf u}}_{M_{0}n,\bar{\alpha}} \right])=c_{n,\bar{\alpha}}{\bf \bar{H}}^{\dagger}{\bf z}_{n,\bar{\alpha}}},
\end{align}
where {${\bf \bar{H}} = \left[\begin{array}{cccc}
{\bf \bar{H}}_{1}^{T} & {\bf \bar{H}}_{2}^{T} & \ldots & {\bf \bar{H}}_{N_{0}}^{T}\end{array}\right]^{T}$}, $c_{n,\bar{\alpha}}$ is a normalization factor, ${\bf z}_{n,\bar{\alpha}}=\left[\begin{array}{ccccccc}
{\bf 0}^{T} & \ldots & {\bf 0}^{T} & {\bf s}_{\bar{\alpha}} & {\bf 0}^{T} & \ldots & {\bf 0}^{T}\end{array}\right]^{T}$, in which {${\bf s}_{\bar{\alpha}}$ is the $n$-th vector of ${\bf z}_{n,\bar{\alpha}}$ and} ${\bf s}_{\bar{\alpha}}=\left[\begin{array}{ccccccc}
0 & \ldots & 0 & 1 & 0 & \ldots & 0\end{array}\right]^{T}\in\mathbb{R}_+^{\left(2L-1\right)\times1}$ (the 1 is located at the $\bar{\alpha}^\text{th}$ index), and ${\left(\cdot \right)^{\dagger}}$
denotes Moore-Penrose pseudo-inverse operator. 

 {
Thus, the $\bar{\alpha}^\text{th}$ candidate for the beamformer component given in \eqref{eq:x_n0} can be represented as
\begin{align}\label{eq:VectorW}
\left[{\bar{\bf u}}_{1n,\bar{\alpha}} \quad {\bar{\bf u}}_{2n,\bar{\alpha}} \quad  \ldots \quad { \bar{\bf u}}_{M_{0}n,\bar{\alpha}} \right]= \text{vec}^{-1}( c_{n,\bar{\alpha}}{\bf \bar{H}}^{\dagger}{\bf z}_{n,\bar{\alpha}} ),
\end{align}
}

{
Specifically,  in the case that the matrix ${\bf \bar{H}}$ has a right-inverse, the inteference components are completely cancelled. Hence, the received signal at  $n^\text{th}$ MU can be simplified as
\begin{align}
{\bf y}_{n}^{0} &= \sqrt{p_{n}^{0}}{ c}_{n,\bar{\alpha}}{\bf s}_{\bar{\alpha}}s_{n}^{0}+\sum\limits _{j=1}^{N_{1}}\sum\limits _{i=1}^{M_{1}}\sqrt{p_{j}^{1}}{\bf g}_{ij}*{\bf h}_{in}^{10}s_{j}^{1} +{\bf n}_{M},
\end{align}
}

On the other hand, for mathematical simplification, we define a factor $\Gamma_{n,\bar{\alpha}}$
shown in \eqref{eq:Gamma_nalpha}. This factor is to evaluate the ratio between the power of the main tap and that of the interferences.
To this end, we present the details of the beamforming design in \textit{Algorithm} 1 for a comprehensive idea.
\begin{center}
\begin{tabular}{|l|}
\hline
\textbf{Algorithm 1}: Algorithm to solve ${\bf u}_{mn}$ \tabularnewline
\hline
\hline
(i). Set $\bar{\alpha}=1$.\tabularnewline
(ii). \textbf{Loop}\tabularnewline
$\quad$1.$\;$Compute ${\bf \bar{u}}_{mn,\bar{\alpha}}$ by \eqref{eq:VectorW1}.\tabularnewline
$\quad$2.$\;$Calculate $\Gamma_{n,\bar{\alpha}}$ by \eqref{eq:Gamma_nalpha}.\tabularnewline
$\quad$3.$\;$Update $\bar{\alpha}\leftarrow\bar{\alpha}+1$.\tabularnewline
$\quad$\textbf{Until} $\bar{\alpha}=2L-1$.\tabularnewline
(iii). Find $\alpha$ with \tabularnewline
$\qquad { \alpha =\arg\max\limits _{\bar{\alpha} }\{\{\Gamma_{n,\bar{\alpha}}\}_{\bar{\alpha}=1}^{2L-1}\}.}$\tabularnewline
(iv). The chosen beamformer ${\bf u}_{mn}$ can be inferred from\tabularnewline
$\quad\quad$${\bf \bar{u}}_{mn,{\alpha}}$.\tabularnewline
\hline
\end{tabular}
\end{center}

\begin{figure*}
\begin{align}\label{eq:Gamma_nalpha}
\Gamma_{n,\bar{\alpha}}=\frac{\left|\left(\sum\limits _{m=1}^{M_{0}}{\bar{\bf u}}_{mn,\bar{\alpha}}*{\bf h}_{mn}^{0}\right)\left[\bar{\alpha}\right]\right|^{2}}{\sum\limits _{l\ne\bar{\alpha}}^{2L-1}\left|\left(\sum\limits _{m=1}^{M_{0}}{\bar{\bf u}}_{mn,\bar{\alpha}}*{\bf h}_{mn}^{0}\right)\left[l\right]\right|^{2}+\sum\limits _{n'\ne n}^{N_{0}}\sum\limits _{l=1}^{2L-1}\left|\left(\sum\limits _{m=1}^{M_{0}}{\bar{\bf u}}_{mn,\bar{\alpha}}*{\bf h}_{mn'}^{0}\right)\left[l\right]\right|^{2}+1}.
\end{align}
\end{figure*}

\subsection{Time Reversal beamforming technique for the FBS}
Unlike MBS, FBS is a low-power cellular station with limited hardware resources \cite{Zah2013,Andrews2012}. This limitation is due to the fact that the zero-forcing beamformer includes the component of matrix inversion, with a huge computational burden that becomes extremely heavy in cases of many users and lengthy CIRs. Furthermore, the transmit power level of FBS is restricted \cite{Zah2013}. Therefore, the zero-forcing technique might not be an interesting solution for femtocell networks. 

In this paper, we propose employing the TR technique to FBS to achieve a better system performance with a much reduced cost.
Indeed, the location signature-specific property of the \gls{tr} can be utilized to mitigate the ISI, the co-tier interference and the cross-tier interference to the macrocell \cite{Bou2013,Eunchul2015}.
 
{
According to \eqref{eq:x_j1}, the FBS beamformer can be expressed by
\begin{align}
\left[\begin{array}{ccc}
                 \gv_{{1}j} & \ldots & \gv_{M_{1}j}
      \end{array} \right] = \left[ {\begin{array}{*{20}{c}}
{{g_{1j}}[1]}& \ldots &{{g_{{M_0}j}}[1]}\\
{{g_{1j}}[2]}& \ldots &{{g_{{M_0}j}}[2]}\\
 \vdots & \vdots & \vdots\\
{{g_{1j}}[L]}& \ldots &{{g_{{M_0}j}}[L]}
\end{array}} \right].
\end{align}}
In TR principle, the time-reversed form of \gls{cir} is employed as \gls{fbs} beamformer, thus each element of $\mathbf{g}_{ij}$ can be calculated as
\begin{align}\label{eq:trbeam}
    {g}_{ij}[l]=\frac{{{h}_{ij}^{1H}[L+1-l]}}{\sqrt{\sum\limits _{i=1}^{M_{1}}\left\Vert {\bf h}_{ij}^{1}\right\Vert ^{2}}}.
\end{align}
Benefiting from the signal focalization property of the TR technique, FU receivers need only to select the central tap to take a sample, i.e. $\beta=L$. As one can evaluate, the TR technique has a much lower computational complexity in comparison to the zero-forcing one.

\section{Proposed Power Allocation Approach}\label{sec:DistributedPowerAllocation}
In this section, we propose a novel downlink power allocation scheme for the considered HetNet taking into account the fact that the backhaul connection is unable to convey all user CSI from femtocell to macrocell. 
We start with briefly presenting the centralized power allocation approach in order to understand this concept and compare its performance to the proposed approach.
In the centralized method, a central controller, likely MBS, {is responsible of computing} the beamforming and power allocation vectors for all \gls{bs}s in the HetNet. 
Assuming that all the CSI of MUs and FUs are available at the MBS, the power control problem which minimizes the total transmit power of BSs {with \gls{sinr} constraints} can be formulated as
\begin{align}\label{eq:MinizeProblem}
\begin{aligned}
	{\text{OP$_{0}:$}\underset{\bf p^{0}, \bf p^{1} }\min} \quad & \sum_{k=0}^{1} \sum_{r=1}^{N_k}p_r^k
\\
	\text{s.t.~} \quad & {\rm SINR}_{r}^{k} \ge\gamma_{r}^{k}, (~\forall k \in \{0,1\};1\le r\le N_{k}),
\end{aligned}
\end{align}
herein $\gamma_r^k$ is the preset threshold for the $r^\text{th}$ user of $\mathcal{B}_k$. Since we divide the downlink power allocation and the beamforming procedures into distinct processes, the optimization problem in \eqref{eq:MinizeProblem} becomes convex on ${\bf p}^{k}$ and the optimal solution can be conveniently found \cite{Ben2001, Cha2009}.

Given this centralized approach, the femtocell needs to send all FU CSI to the macrocell via the backhaul link. Therefore, signaling overheads as well as computational burdens at the macrocell are heavy when the network size is large.
However, in case of {link congestion}, obtaining sufficient \gls{cir} becomes intractable.
Due to these drawbacks, a novel scheme is proposed for the considered HetNet. In this vein, the original optimization problem OP$_0$ given in equation \eqref{eq:MinizeProblem} is decomposed into two subproblems, i.e.  MBS subproblem and FBS subproblem, to (i) share computational burden to all BSs and (ii) reduce the dependence on backhaul links. 
Different from previous works \cite{Tolli2011, Joe2011, Duy2015}, our proposed optimization method can perform as follows: (i) To reduce the amount of signaling information, the FBS computes first the beamforming vector for its own users, then optimizes the corresponding transmission power to mitigate the cross-tier interference to the MUs and finally transfers the minimized value of the cross-tier interference power to the \gls{mbs}. (ii) Afterwards, the MBS initiates computing its own beamformer and power allocation vector once it receives the specific information sent by the FBS. Hence, this method limits the signaling overhead since only a minimized value of the cross-tier interference power is required at the MBS side. In addition, the FUs are released from the task of measuring the cross-interference caused by the MBS.
Importantly, our scheme can allow the operator to control the priority of MUs, shown in subsection IV.A.
Furthermore, the optimal closed-form solutions are derived for such subproblems. A comparison between two the centralized and proposed approaches is shown in Fig. 2.
In this regard, the required amount of signaling exchange between two tiers becomes much smaller (i.e. Table I), and the proposed scheme can achieve near optimal performance.
To this end, the subproblems for femtocell and macrocell are thoroughly discussed in the following subsections.

\begin{figure}[!]
\includegraphics[scale=0.35]{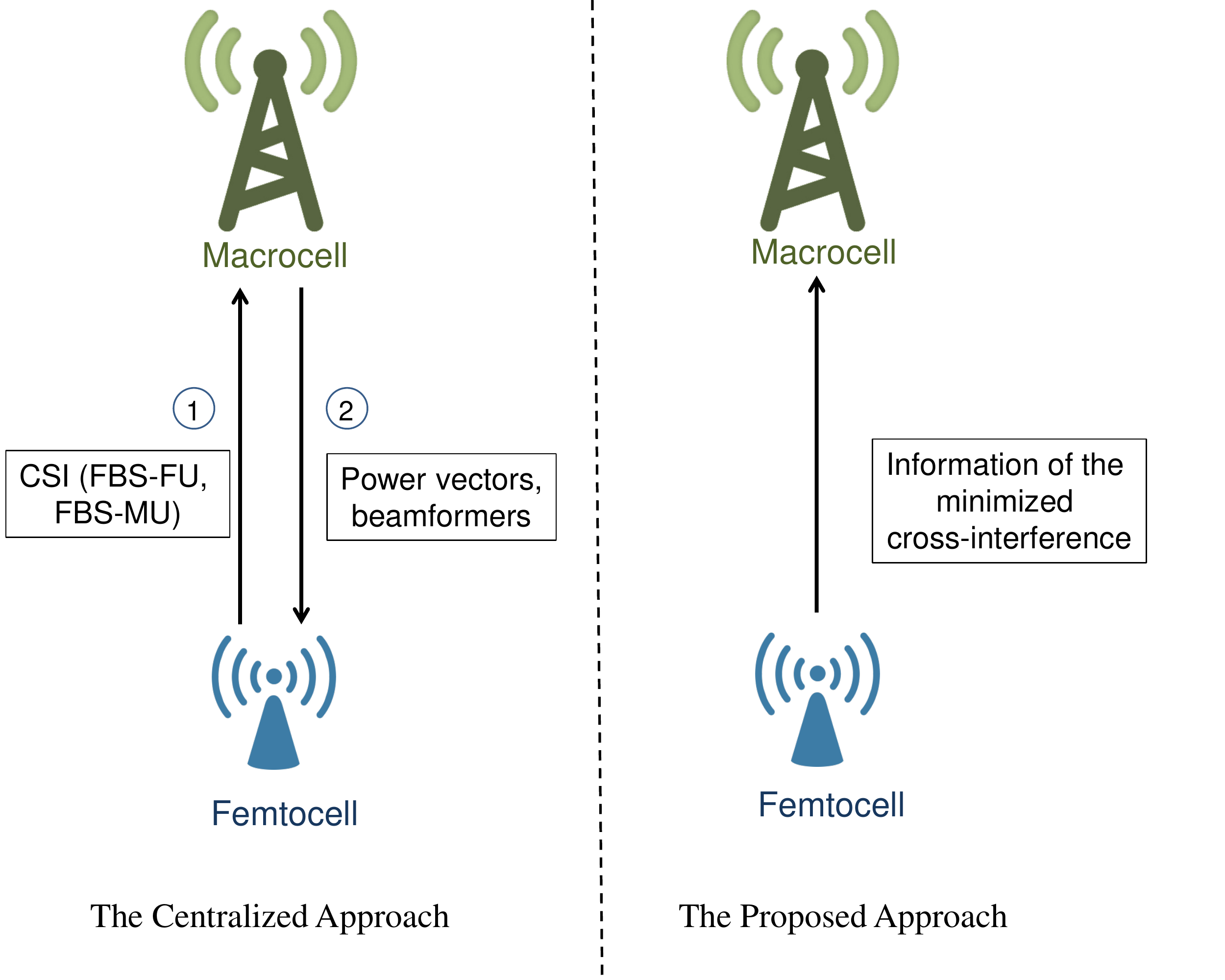}
\caption{A comparsion between the approaches.}
\label{fig:scheme}
\end{figure}

\begin{table}
\begin{centering}
\caption{The CSI required at MBS}
\par\end{centering}
\centering{}%
\begin{tabular}{|l|c|c|}
\hline
\textbf{Kinds of CSI} &
\textbf{Centralized approach}&
\textbf{Proposed approach}
\tabularnewline
\hline
MBS - MU & required & required
\tabularnewline
\hline
MBS - FU & required & required
\tabularnewline
\hline
FBS - MU & required & not required
\tabularnewline
\hline
FBS - FU & required & not required
\tabularnewline
\hline
\end{tabular}
\end{table}

\subsection{Subproblem 1: Power allocation for the FBS}
In HetNet, both \glspl{mu} and \glspl{fu} frequently endure the cross-tier interference,
but \glspl{mu} have a higher priority than FUs in communication such that optimization designs must satisfy \glspl{mu} \gls{qos}. Therefore, to give priority to \glspl{mu}, the femtocell should minimize the interference that it causes to \glspl{mu}. On the other hand, to reduce the CSI sent to MBS, the femtocell starts the communication by providing an optimized power allocation with respect to a tolerable level of cross-tier interference that is set for each FU as $P_{(tol)}{}_{j}^{01}$. 
Accordingly, the downlink power control problem for the \gls{fbs} can be interpreted as
\begin{align}\label{eq:SUB-OP1}
    \begin{aligned}
        {\text{OP$_1$:~} \underset{\bf p^{1}}\min}\quad & \sum\limits _{j=1}^{N_{1}}\left(\sum\limits _{n=1}^{N_{0}}\left\Vert \sum\limits _{i=1}^{M_{1}}{\bf g}_{ij}*{\bf h}_{in}^{10}\right\Vert ^{2}p_{j}^{1}\right)
        \\
        \text{s.t.}\quad & \text{SINR}_{j}^{1}\left(L,\;{\bf p}^{1},\;\left\{ {\bf g}_{ij}\right\} _{i=1,j=1}^{M_{1},N_{1}},\; P_{(tol)}{}_{j}^{01}\right)\ge\gamma_{j}^{1}.\\
	&{(1\le j\le N_{1})}
    \end{aligned}
\end{align}
Hence, the tolerable cross-tier interference implies that
\begin{align}\label{eq:cross}
 {P_{(cross)}{}_{j}^{1}} = \sum\limits _{n=1}^{N_{0}}\left\Vert \sum\limits _{m=1}^{M_{0}}\sqrt{p_{n}^{0}}{\bf u}_{mn}*{\bf h}_{mj}^{01}\right\Vert ^{2} \le P_{(tol)}{}_{j}^{01}.
\end{align} 
Moreover, the inequality given in \eqref{eq:cross} is set as a constraint of the MBS's power allocation problem, i.e. see  \eqref{eq:OP2}. 

In fact, although MUs are primary users, there should be restrictions applied in the cross-interference that MBS cause to FUs due to flexible management for operators. Given this scheme, the operator might manage the priority level of MUs by adjusting the tolerable threshold, i.e. $ \{P_{(tol)}{}_{j}^{01}\}$. This makes the proposed scheme more flexible than the previous works \cite{Joe2011, Duy2015}.

It is easy to see that the considered optimization problem requires the \gls{csi} at femtocells only, and the objective function aims to minimize the cross interference to MUs, i.e. $ {P_{(cross)}{}_{n}^{0}}$. 
On the other hand, it is visible that the problem OP$_1$ is a linear programming problem. Thus, the solution of problem OP$_1$ is summarized in \emph{Lemma} 1 below.

\newtheorem{lem}{Lemma}
\begin{lem}
Let ${\bf p}^{\star 1}$ denote the optimal value of ${\bf p}^{1}$. Based on preliminaries in {[18.4 \cite{Ben2001}],  [eq. (4) \cite{Cha2009}],} the closed-form expression of
${\bf p}^{\star 1}$ can be given by
\begin{align}\label{eq:p_op}
{\bf p}^{\star 1}={\rm diag}({\pmb \eta})^{-1}\left[{\bf I}_{N_{1}}-{\rm diag}({\pmb \eta} ){\bf D}\left({\bf B}\circ ({\pmb \eta}^{-1} {\bf 1}_{N_{1}\times {1}}^T )\right) \right]^{-1}{\bf z},
\end{align}
where $\circ$ denotes Hadamard product, ${\bf I}_{N_{1}}$ is an $N_{1}\times N_{1}$ identity matrix, \textbf{$\mathbf{B}$}
is a $N_{1}\times N_{1}$ matrix whose the $(j,\; j')^\text{th}$ entry is
defined as
\begin{align}
\mathbf{(B)}_{j'j}=\begin{cases}
0, & j=j'\\
{\left\Vert \sum\limits _{i=1}^{M_{1}}{\bf g}_{ij'}*{\bf h}_{ij}^{1}\right\Vert ^{2}}, & j\neq j',
\end{cases}
\end{align}
in addition,
\begin{align}
{\pmb \eta} = \dfrac{\hat{\pmb \eta}}{\left\| {\hat{\pmb \eta}} \right\|},
\end{align}
in which each element of the vector $\hat {\pmb \eta} \in \mathbb C^{N_1}$ is

\begin{align}
\hat \eta_{j}&=\sum\limits _{n=1}^{N_{0}}\left\Vert \sum\limits _{i=1}^{M_{k}}{\bf g}_{ij}^{}*{\bf h}_{in}^{10}\right\Vert ^{2},
\end{align}
and $\mathbf{D}$ is a $N_{1}\times N_{1}$ matrix which is presented by
\begin{align}
{\bf D}={\rm diag}\left\{ \frac{\gamma_{1}^{1}}{\varphi_{1}};\;\ldots;\;\frac{\gamma_{N_{1}}^{1}}{\varphi_{N_{1}}}\right\},
\end{align}
where
\begin{align}
\varphi_{j}={ \left|\left(\sum\limits _{i=1}^{M_{1}}{\bf g}_{ij}*{\bf h}_{ij}^1\right)\left[L\right]\right|^{2}-\gamma_{j}^{1}\sum\limits _{{l=1\atop l\ne L}}^{2L-1}\left|\left(\sum\limits _{i=1}^{M_{1}}{\bf g}_{ij}*{\bf h}_{ij}^{1}\right)\left[l\right]\right|^{2}   },
\end{align}
and $\textbf z$ is a vector $\textbf z=[z_1~z_2~\ldots~z_{N_1}]^T$ with each element given as
\begin{align}
{z}_j=
P_{(tol)j}^{01}+\left\Vert {\bf n}_{F}[L]\right\Vert ^{2}.
\end{align}
\end{lem}

\begin{IEEEproof}
See Appendix A.
\end{IEEEproof}

In the proposed scheme, when the femtocell tackles the optimization problem OP$_{1}$, the value of $\left\{P^\star_{(cross)}{}_{n}^{0}\right\}_{n=1}^{N_0}$ given in equation \eqref{eq:P_cross_min} is sent to the macrocell instead of the information of ${\bf p}^{\star 1}$, $\left\{ {\bf h}_{ij}^1\right\} _{i=1,\: j=1}^{M_{1},\: N_{1}}$
and $\left\{ {\bf h}_{in}^{10}\right\} _{j=1,\: n=1}^{N_{1},\: N_{0}}$  via the backhaul link. 
\begin{align} \label{eq:P_cross_min}     
{P^\star_{(cross)}{}_{n}^{0}}&=\sum\limits _{j=1}^{N_{1}}\left\Vert \sum\limits _{i=1}^{M_{1}}\sqrt{p_{j}^{\star1}}{\bf g}_{ij}*{\bf h}_{in}^{10}\right\Vert ^{2}.
\end{align}
{
It is worth reminding that ${\bf h}_{in}^{10}$ denotes the CIR between the femtocell and the $n^\text{th}$ macrocell user.
In other words, ${\bf h}_{in}^{10}$ is the local CSI of the femtocell network. In our paper, ${\bf h}_{in}^{10}$ is assumed to be available at the femtocell in both the centralized and the proposed approaches.
}

{
Therefore, in the proposed power allocation, one can conclude that the amount of overhead used for signaling information is significantly reduced.}

\subsection{Subproblem 2: Power loading problem for the MBS}

In this part, we present our derivation methodology of the downlink power allocation for MBS. Our aim is to minimize the total transmit power with the interference constraint to FUs. 
To actively guarantee the performance for \glspl{mu}, the \gls{mbs} computes the beamforming and power allocation vectors once it receives the signaling information of $\left\{P^\star_{(cross)}{}_{n}^{0}\right\}_{n=1}^{N_0}$ given by \eqref{eq:P_cross_min} from the femtocell.
Hence, the optimization problem involving the \gls{sinr} and interference constraints is formulated as
\begin{align}\label{eq:OP2}
    \begin{aligned}
    {\text{ OP$_2$:~}} &\underset{\mathbf{{\bf p}^{0}}} \min \quad \sum\limits _{n=1}^{N_{0}} {p_n^0}
    \\
    \text{s.t.}\quad & {\rm SINR}_{n}^{0}\left(\alpha,{\bf p}^{0},\left\{ {\bf u}_{mn}\right\} _{m=1,n=1}^{M_{0},N_{0}},\; P^\star_{(cross)}{}_{n}^{0}\right)\ge\gamma_{n}^{0}.
\\
&\left\Vert \sum\limits _{m=1}^{M_{0}}\sqrt{p_{n}^{0}}{\bf u}_{mn}*{\bf h}_{mj}^{01}\right\Vert ^{2}\le P_{(tol)}{}_{j}^{01}, 
    \end{aligned}
\end{align}
The above problem is a linear programming problem which can be solved by interior-point method. To reduce the
computational burden, we aim at solving OP$_2$ by closed-form expressions. Since the objective function and the constrains of OP$_2$ are not differentiable, it is infeasible to solve OP$_2$ through its the Lagrangian dual. To deal with this issue, we endeavour to transform the problem OP$_2$ into an equivalent formulation solvable by the Lagrange multiplier method.

In this context, we start with applying the uplink-downlink duality property to the \gls{sinr}$^0_n$ constraint of the OP$_2$. It is observed that the \gls{sinr}$^0_n$ constraint can be considered as a function of $P_{(co)}{}_{n}^{0}$.
According to the property of uplink-downlink duality, the virtual uplink \gls{sinr}$^0_n$ derivation denoted by $\overline{\rm SINR}_{n}^{0}$ has the same structure as \gls{sinr}$^0_n$ expression, however, $P_{(co)}{}_{n}^{0}$ is replaced by $\overline{P}_{(co)}{}_{n}^{0}$, with
\begin{align}
\overline{P}_{(co)}{}_{n}^{0}=\sum\limits _{{n'=1\atop n'\ne n}}^{N_{0}}\left\Vert \sum\limits _{m=1}^{M_{0}}\sqrt{p_{n}^{0}}{\bf u}_{mn}*{\bf h}_{mn'}^{0}\right\Vert ^{2}.
\end{align}
\begin{figure*}
\begin{align}\label{eq:Delta_n}
 &\Delta_{n}=\frac{\sum\limits _{l\ne\alpha}^{2L-1}\left|\left(\sum\limits _{m=1}^{M_{0}}{\bf u}_{mn}*{\bf h}_{mn}^{0}\right)\left[l\right]\right|^{2}+\sum\limits _{{n'=1\atop n'\ne n}}^{N_{0}}\left\Vert \sum\limits _{m=1}^{M_{0}}{\bf u}_{mn}*{\bf h}_{mn'}^{0}\right\Vert ^{2}}{\left|\left(\sum\limits _{m=1}^{M_{0}}{\bf u}_{mn}*{\bf h}_{mn}^{0}\right)\left[\alpha\right]\right|^{2}}.
\end{align}
\end{figure*}

Since ${\bf p}_{n}^{0}$ is a non-negative vector, it can be re-written in the new form
\begin{align}
\exp\left(\xi_{n}\right)=p_{n}^{0}.
\end{align}
On this basis, the inversion of $\overline{\rm SINR}_{n}^{0}$ can be computed as
\begin{align}
\left(\overline{\rm SINR}_{n}^{0}\right)^{-1}=\Delta_{n}+\nabla_{n}\exp\left(-\xi_{n}\right),
\end{align}
where $\Delta_{n}$ is defined as in \eqref{eq:Delta_n} and
\begin{align}
\nabla_{n}=\frac{P^\star_{(cross)}{}_{n}^{0}+\left\Vert {\bf n}_{M}\right\Vert ^{2}}{\left|\left(\sum\limits _{m=1}^{M_{0}}{\bf u}_{mn}*{\bf h}_{mn}^{0}\right)\left[\alpha\right]\right|^{2}}.
\end{align}
Without loss of generality, the problem OP$_2$ is reformulated into a more tractable formulation such as
\begin{align}
    \begin{aligned}
{\text{OP$_{3}$:}}& \underset{\{{\xi_{n}}\}_{n=1}^{N_0}}{\text{min}}\quad \sum\limits _{n=1}^{N_0}\exp\left(\xi_{n}\right)\\
\text{s.t.}\quad &\log\left(\Delta_{n}+\nabla_{n}\exp\left(-\xi_{n}\right)\right)\le\log\frac{1}{\gamma_{n}^{0}},\\
&\left\Vert \sum\limits _{m=1}^{M_{0}}{\bf u}_{mn}*{\bf h}_{mj}^{01}\right\Vert ^{2}\exp\left(\xi_{n}\right)\le P_{(tol)}{}_{j}^{01}.
    \end{aligned}
\end{align}
Indeed, the problem OP$_{3}$ is convex and it is solvable by use of Lagrange multiplier method \cite{Gra2009}. Considering the relationship between OP$_{2}$ and OP$_{3}$, it is clear that 
the optimal solution of OP$_{2}$ can be calculated according to that of OP$_{3}$, i.e. ${p}_{n}^{\star 0}=\exp\left({\xi}_{n}^\star\right)$ where ${\xi}_{n}^\star$ is the optimal solution of OP$_{3}$.

In the continuity, following Lagrange multiplier method, let $\left\{ \mu_{n}\right\}$ and $\left\{ \lambda_{n}\right\}$ be the dual variables associated with the SINR and interference constraints, respectively.
After some mathematical manipulations, the Lagrangian function of OP$_{3}$ can be given by
\begin{align}\label{eq:Lag1}
 &\mathbb{\mathscr{L}}\left(\left\{ \xi_{n}\right\} ,\left\{ \mu_{n}\right\} ,\left\{ \lambda_{n}\right\} \right)
  =\sum\limits _{n=1}^{N_0}\exp\left(\xi_{n}\right)\nonumber
  \\
  &\hspace{+1cm}+\sum\limits _{n}^{N_0}\mu_{n}\left(\log\left(\Delta_{n}+\nabla_{n}\exp\left(-\xi_{n}\right)\right)-\log\frac{1}{\gamma_{n}^{0}}\right)
  \nonumber
  \\  
 & \hspace{+1cm}+\sum\limits _{n=1}^{N_0}\lambda_{n}\left(\left\Vert \sum\limits _{m=1}^{M_{0}}{\bf u}_{mn}*{\bf h}_{mj}^{01}\right\Vert ^{2}\exp\left(\xi_{n}\right)-P_{(tol)}{}_{j}^{01}\right).
\end{align}
For convenience, we focus on the following simplified form of \eqref{eq:Lag1}
\begin{align}\mathscr{L}_{n}\left(\xi_{n},\mu_{n},\lambda_{n}\right)&= \exp\left(\xi_{n}\right)+\mu_{n}\log\left(\Delta_{n}+\nabla_{n}\exp\left(-\xi_{n}\right)\right)
\nonumber
\\
 &+\lambda_{n}\left\Vert \sum\limits _{m=1}^{M_{0}}{\bf u}_{mn}*{\bf h}_{mj}^{01}\right\Vert ^{2}\exp\left(\xi_{n}\right).
\end{align}
Accordingly, the dual function and the dual problem can explicitly be formulated as \eqref{eq:D1} and \eqref{eq:D2}, respectively
\begin{align}\label{eq:D1}
&\mathscr{D}\left(\left\{ \xi_{n}\right\} ,\left\{ \mu_{n}\right\} ,\left\{ \lambda_{n}\right\} \right)
=\sum\limits _{n=1}^{N_0}\min\limits _{\xi_{n}}\mathscr{L}_{n}\left(\xi_{n},\mu_{n},\lambda_{n}\right) 
\\
 &-\sum\limits _{n=1}^{N_0}\mu_{n}\left(\log\frac{1}{\gamma_{n}^{0}}\right)
 -\sum\limits _{n=1}^{N_0}\lambda_{n}\left(P_{(tol)}{}_{j}^{01}\right),\nonumber
\end{align}
\begin{align}\label{eq:D2}
\underset{\left\{ \mu_{n}\right\} ,\left\{ \lambda_{n}\right\} }{\text{OP$_4$:}}~\max & \mathscr{\quad D}\left(\left\{ \mu_{n}\right\} ,\left\{ \lambda_{n}\right\} \right)
\\
\text{s.t.} & \quad \left\{\mu_{n}\right\}\ge0,\;\left\{\lambda_{n}\right\}\ge0.\;\forall n \nonumber
\end{align}

It is a fact that the problem OP$_3$ is a convex optimization problem, thus strong duality holds, i.e. the duality gap between
primal problem and dual problem is zero. According to Karush-Kuhn-Tucker condition, the optimal transmit power can be obtained through the first derivative of the Lagrangian function with respect to $\xi_{n}$ as
\begin{align}
    \frac{\partial\mathscr{L}_{n}\left(\xi_{n},\mu_{n},\lambda_{n}\right)}{\partial\xi_{n}}=0.
\end{align}

Specifically, the optimal solution ${p}_{n}^{\star 0}$ can be achieved by the closed-form derivation
\begin{align} {p}_{n}^{\star 0}&=\exp\left({\xi}_{n}^{\star}\right) \nonumber
\\
 & =\frac{2\mu_{n}\nabla_{n} \lambda_{n}^{-1}\left\Vert \sum\limits _{m=1}^{M_{0}}{\bf u}_{mn}*{\bf h}_{mj}^{01}\right\Vert ^{-2} }{\left(\nabla_{n}+\sqrt{\nabla_{n}^{2}+\dfrac{4\nabla_{n}\Delta_{n}\mu_{n}}{\lambda_{n}\left\Vert \sum\limits _{m=1}^{M_{0}}{\bf u}_{mn}*{\bf h}_{mj}^{01}\right\Vert ^{2}}}\right)}.
\end{align}
The final step in this method is to provide optimal Lagrangian multipliers, i.e. $\mu_n$ and $\lambda_{n}$ by solving the dual problem OP$_4$. In this way, the optimal solution of OP$_3$ is obtained once $\exp\left(\xi_{n}\right)$, $\mu_n$ and $\lambda_{n}$ are iteratively updated until convergence.
The subgradient iteration algorithm is then applied to update the Lagrangian multipliers as 
\begin{align} 
\mu_{n}\left(t\right)&=\left[\mu_{n}\left(t-1\right)+\nu_{n}\left(t\right) X_1 \right]^{+},
\\
\lambda_{n}\left(t\right)
 &=\left[\lambda_{n}\left(t-1\right)+\kappa_{n}\left(t\right)X_2\right]^{+},
\end{align}
where $\nu_{n}\left(t\right)$ and $\kappa_{n}\left(t\right)$
are step sizes, while $X_1$ and $X_2$ are defined as
\begin{align}
    X_1&=\log\left(\Delta_{n}+\nabla_{n}\exp\left(-\xi_{n}\right)\right)-\log\frac{1}{\gamma_{n}^{0}},
    \\
    X_2&=\left\Vert \sum\limits _{m=1}^{M_{0}}{\bf u}_{mn}*{\bf h}_{mj}^{01}\right\Vert ^{2}\exp\left(\xi_{n}\right)-P_{(tol)}{}_{j}^{01}.
\end{align}
The subgradient iterative algorithm is ensured to converge to the optimal value with a sufficiently small step size \cite{Gra2009}.

\section{Worst-case Robust Optimization for TR Femtocell Network}\label{sec:WorstCase}
\begin{figure*}
\begin{align}\label{eq:WorstCaseOptimum2}
 &\min\limits _{{\bf e}_{ij}^{1},{\bf e}_{ij'}^{1}} \text{SINR}_{j}^{1}\left(\beta,{\bf \; p}^{1},\;\left\{ \hat {\bf g}_{ij}^{1}\right\} _{i=1,j=1}^{M_{1},N_{1}}, P_{(tol)}{}_{j}^{01}\right)
 \approx\frac{\mathcal Pl_{(sig)}{}_{j}^{1}\left(p_{j}^{1},{\bf \hat{g}}_{ij}^{}\right)}{\mathcal Pu_{(isi)}{}_{j}^{1}\left(p_{j}^{1},{\bf \hat{g}}_{ij}^{}\right){\rm +} \mathcal Pu_{(co)}{}_{j}^{1}\left(\{p_{j'}^{1},{\bf \hat{g}}_{ij'}^{}\}_{j'=1,j'\ne j}^{N_{1}}\right)+P_{(tol)}{}_{j}^{01}+\left\Vert {\bf n}_{F}[L]\right\Vert ^{2}}.
\end{align}
\end{figure*}
In this section, we extend the downlink power minimization problem for the TR femtocell by considering imperfect CSI. It is worth noting that this work considers for the first time the worst-case robust optimization for TR femtocell network.

Accounting for error model, we assume that the maximum error quantity of the femtocell can be known, such that the imperfect \gls{cir} model can be referred to as
\begin{align}\label{eq:error}
{\bf \hat{h}}_{ij}^{1}={\bf h}_{ij}^{1}+{\bf e}_{ij}^{1},
\end{align}
where ${\bf e}_{ij}^{1}$ represents a channel uncertainty defined by a feasible set as
\begin{align}\label{eq:error1}
\mathcal F_{ij}^1 = \left\{{\bf e}_{ij}^{1} \in \mathbb C^{L \times 1}:  \left\Vert {\bf e}_{ij}^{1}\right\Vert ^{2}\le\psi\left\Vert {\bf h}_{ij}^{1}\right\Vert ^{2} \right\}, 
\end{align}
and $\psi$ is named as error factor.
We also suppose that ${\bf h}_{ij}^{1}$ and ${\bf e}_{ij}^{1}$ are identically and independently distributed
variables.
Under the influence of CEE, we devise the robust optimization methodologies to guarantee the QoS requirement. Following the worst-case approach, the power allocation design of the femtocells may be formulated as

\begin{align}\label{eq:WorstCaseOptimum1}
&\text{OP$_5$:~} \underset{\mathbf{p}^{1}}{\text{min}} \quad\max\limits _{{\bf e}_{in}^{10}}\sum\limits _{j=1}^{N_{1}}\left(\sum\limits _{n=1}^{N_{0}}\left\Vert \sum\limits _{i=1}^{M_{1}}{\bf \hat{g}}_{ij}^{}*{\bf h}_{in}^{10}\right\Vert ^{2}p_{j}^{1}\right)
\\
&\text{s.t.} \min\limits _{{\bf e}_{ij}^{1},{\bf e}_{ij'}^{1}} \text{SINR}_{j}^{1}\left(L,{\bf \; p}^{1},\;\left\{ \hat {\bf g}_{ij}^{1}\right\} _{i=1,j=1}^{M_{1},N_{1}}, P_{(tol)}{}_{j}^{01}\right) \ge\gamma_{j}^{1}
\nonumber \\
& \quad \quad \quad \left( {{\bf e}_{in}^{10}} \in \mathcal F_{in}^{10}, {{\bf e}_{ij}^{1}} \in \mathcal F_{ij}^{1}, {{\bf e}_{ij'}^{1}} \in \mathcal F_{ij'}^{1} \right). \nonumber
\end{align}
Unluckily, the problem OP$_5$ is intractable since the objective function and constraint include the convolution operator. However, it can be transformed into a convex one by approximating the latters.
Indeed, one can see that it is challenging to obtain an exact closed-form \gls{sinr} expression in the constraint.

To tackle this problem, we apply an approximation presented as \eqref{eq:WorstCaseOptimum2} in which ${ {\mathcal Pl}_{(sig)}}_j^1 $, ${ {\mathcal Pu}_{(isi)}}_j^1$ and ${ {\mathcal Pu}_{(co)}}_j^1$ are the worst-case lower-bound of central signal power, the worst-case upper-bound of ISI power, and worst-case upper-bound of co-tier interference power, respectively.
\begin{align}\label{eq:Plsig}
{{\mathcal Pl}_{(sig)}}_j^1 &= \underset{{\bf e}_{ij}^{1} \in \mathcal F_{ij}^1}{\min} \left|\sum\limits _{i=1}^{M_{1}}({\hat{\bf g}}_{ij}^{}*{ {\bf h}}_{ij}^{1})[L]\sqrt{p_{j}^{1}}\right|^{2},
\end{align}
\begin{align}\label{eq:Plisi}
{ {\mathcal Pu}_{(isi)}}_j^1 &= \underset{{\bf e}_{ij}^{1} \in \mathcal F_{ij}^1}{\max} \left\Vert \sum\limits _{i=1}^{M_{1}}{\bf \hat{g}}_{ij}^{}*{\bf h}_{ij}^{1}\right\Vert ^{2} - { {\mathcal Pl}_{(sig)}}_j^1,
\end{align}
\begin{align}\label{eq:Plco}
{ {\mathcal Pu}_{(co)}}_j^1 &= \underset{{\bf e}_{ij'}^{1} \in \mathcal F_{ij'}^1}{\max} \sum\limits _{{j'=1\atop j'\ne j}}^{N_{1}}\left\Vert \sum\limits _{i=1}^{M_{1}}\sqrt{p_{j'}^{1}}{\bf {\hat g}}_{ij'}*{\bf h}_{ij}^{1}\right\Vert ^{2},
\end{align}
where ${\bf {\hat g}}_{ij}$ is the beamformer corresponding to the estimated channel ${\bf \hat{h}}_{ij}^{1}$, and it has a similar structure to ${\bf {g}}_{ij}$ shown in \eqref{eq:trbeam}.
Note that the derivation of ${ {\mathcal Pu}_{(isi)}}_j^1$ can be expressed as a substraction between the upper-bound of the power of all taps and the lower-bound of the power of central tap. 
Our aim is to derive such these boundaries into formulations that can be expressed in term of estimated values solely, e.g. ${\hat{\bf g}}_{ij}^{}$, ${\hat {\bf h}}_{ij}^{1}$.

\subsection{Worst-case lower-bound of signal power component}
This part of the paper is dedicated to the derivation of the worst-case lower-bound of signal power given in equation \eqref{eq:Plsig}.

With an erroneous channel estimation, the power of the central tap at the intended user becomes
\begin{align}\label{eq:fol1}
&\left|\sum\limits _{i=1}^{M_{1}} \left({\hat{\bf g}}_{ij}^{}*{\bf h}_{ij}^{1}\right)[L]\right|^{2} \nonumber \\
& =\frac{\left|\sum\limits _{i=1}^{M_{1}} \left( \sum\limits _{l=1}^{L}{ h}_{ij}^{1}[l]{h}_{ij}^{*1}[l]+\sum\limits _{l=1}^{L}{e}_{ij}^{1}[l]{h}_{ij}^{*1}[l] \right) \right|^{2}}{\sum\limits _{i=1}^{M_{1}}\left\Vert {\bf \hat{h}}_{ij}^{1}\right\Vert ^{2}} \nonumber
\\
 & =\frac{\left| \sum\limits _{i=1}^{M_{1}} \left( \left\Vert {\bf h}_{ij}^{1}\right\Vert ^{2}+({\bf e}_{ij}^{1})^{H}{\bf h}_{ij}^{1} \right)\right|^{2}}{\sum\limits _{i=1}^{M_{1}}\left\Vert {\bf \hat{h}}_{ij}^{1}\right\Vert ^{2}},
\end{align}
where we have
\begin{align}\label{eq:fol2}
\left\Vert {\bf \hat{h}}_{ij}^{1}\right\Vert ^{2}=\left\Vert {\bf h}_{ij}^{1}\right\Vert ^{2}+\left\Vert {\bf e}_{ij}^{1}\right\Vert ^{2}+2\Re\left\{({\bf e}_{ij}^{1})^{H}{\bf h}_{ij}^{1}\right\}.
\end{align}
One can evaluate that while $\left\Vert {\bf \hat{h}}_{ij}^{1}\right\Vert ^{2}$ is fixed,
if $\left\Vert {\bf e}_{ij}^{1}\right\Vert ^{2}$ increases then $\left\Vert {\bf h}_{ij}^{1}\right\Vert ^{2}$
decreases. Considering \eqref{eq:fol1}, the increasing level of $({\bf e}_{ij}^{1})^{H}{\bf h}_{ij}^{1}$
is generally lower than the decreasing level of $\left\Vert {\bf h}_{ij}^{1}\right\Vert ^{2}$.
Therefore, the CEE effect  monotonically reduces the power of desired signal, and the power focalization is decreased. \emph{Lemma} 2 below determines ${\mathcal Pl}_{(sig)}{}_{j}^{1}$ based on \eqref{eq:fol1}.
\begin{lem}[]
The worst-case lower-bound on the signal power component can be stated as
\begin{align}\label{eq:Optimum1}
{\mathcal Pl}_{(sig)}{}_{j}^{1}=\left|\sum\limits _{i=1}^{M_{1}}({\hat{\bf g}}_{ij}^{}*{ \hat{\bf h}}_{ij}^{1})[L]\right|^{2} \frac{{p_{j}^{1}}}{\left(1-\sqrt{\psi}\right)^{2}}.
\end{align} 
\end{lem}
\begin{IEEEproof}
The proof is listed in Appendix B.
\end{IEEEproof}

\subsection{Worst-case upper-bound on the ISI power component}
In order to find out worst-case upper-bound on the \gls{isi} term, the maximum
of Euclidean norm of $\sum\limits _{i=1}^{M_{1}}{\bf \hat{g}}_{ij}^{}*{\bf h}_{ij}^{1}$, i.e. $ \underset{{\bf e}_{ij}^{1} \in \mathcal F_{ij}^1}{\max} \left\Vert \sum\limits _{i=1}^{M_{1}}{\bf \hat{g}}_{ij}^{}*{\bf h}_{ij}^{1}\right\Vert ^{2}$ needs to be discovered first. This problem is non-trivial since only the knowledge of the norm constraint of the estimation error is available, see \eqref{eq:error1}. Fortunately, thanks to  Young\textquoteright{}s inequality \cite{You1912}, {[}eq (3.9.4) \cite{Bog2007}{]} the upper-bound can be derived. In this vein, the norm of convolution between the given two vectors can be bounded by Young\textquoteright{}s inequality as
\begin{align} \label{eq:Bound}
&\left\Vert \sum\limits _{i=1}^{M_{1}}{\bf \hat{g}}_{ij}^{}*{\bf h}_{ij}^{1}\right\Vert ^{2} \le \hat c\sum\limits _{i=1}^{M_{}}\left\Vert {\bf h}_{ij}^{1}\right\Vert ^{2}\left\Vert {\bf \hat{g}}_{ij}^{}\right\Vert _{1}^{2}
\\
 &+ \hat c \sum\limits _{i=1}^{M_{1}}\sum\limits _{{i'=1\atop i'\ne i}}^{M_{1}}\left\Vert {\bf h}_{ij}^{1}\right\Vert \left\Vert {\bf \hat{g}}_{ij}^{}\right\Vert _{1}\left\Vert {\bf h}_{i'j}^{1}\right\Vert \left\Vert {\bf \hat{g}}_{ij}^{}\right\Vert _{1}, \nonumber 
\end{align}
where $\left\Vert \cdot\right\Vert _{1}$ denotes the $l_{1}$\textendash{}norm and $\hat c$ is a constant.
One can see that the resulting boundary can be computed as a function of the norm of these two vectors only. Concerning the worst possible error case, based on the defined feasible set given in \eqref{eq:error1}, we have
\begin{align}
&\underset{{\bf e}_{ij}^{1} \in \mathcal F_{ij}^1}{\max} \left\Vert \sum\limits _{i=1}^{M_{1}}{\bf \hat{g}}_{ij}^{}*{\bf h}_{ij}^{1}\right\Vert ^{2} = \hat c \sum\limits _{i=1}^{M_{1}}\frac{\left\Vert {\bf \hat{h}}_{ij}^{1}\right\Vert ^{2}\left\Vert {\bf \hat{g}}_{ij}^{}\right\Vert _{1}^{2}}{\left(1-\sqrt{\psi}\right)^{2}} \\
&+ \hat c\sum\limits _{i=1}^{M_{1}}\sum\limits _{{i'=1\atop i'\ne i}}^{M_{1}}\frac{\left\Vert {\bf \hat{h}}_{ij}^{1}\right\Vert \left\Vert {\bf \hat{g}}_{ij}^{}\right\Vert _{1}\left\Vert {\bf \hat{h}}_{i'j}^{1}\right\Vert \left\Vert {\bf \hat{g}}_{i'j}^{}\right\Vert _{1}}{\left(1-\sqrt{\psi}\right)^{2}}. \nonumber
\end{align}

On the other hand, the objective function and ${{\mathcal Pu}_{(iui)}}_j^1$ also contain the worst-case boundary of the norm of convolution between two vectors.
It is a fact that the tight degree of the boundary plays an important role in limiting the waste of the transmit power allocation. For a long time, the designation of a value to the constant $\hat c$ in \eqref{eq:Bound} that can improve the tightness of Young's inequality was a challenge for researchers. Eventually, Beckner \cite{Bec1975} provided the best possible constant $\hat c$ and the work in \cite{Sergey2011} generalized Young's inequality, the value of $\hat c$ designed by \cite{Bec1975,Sergey2011} has no effect in the case considered in our work, i.e. $\hat c$ is equal to 1.
This motivates us to derive a tighter worst-case upper-bound on the \gls{isi}
component through \emph{Lemma} 3. 
\begin{lem}
Considering the worst-case boundary of the norm of convolution between two vectors, we introduce a new formulation as follows
\begin{align}
   &\underset{{\bf e}_{ij}^{1} \in \mathcal F_{ij}^1}{\max} \left\Vert \sum\limits _{i=1}^{M_{1}}{\bf \hat{g}}_{ij}^{}*{\bf h}_{ij}^{1}\right\Vert ^{2} \\
    & = \sum\limits _{i=1}^{M_{1}}\left\Vert {\bf \hat{g}}_{ij}^{}*{\bf \vec{h}}_{ij}^{\star 1}\right\Vert ^{2}+\left| \sum\limits _{i=1}^{M_{1}}\sum\limits _{{i=1\atop i'\ne i}}^{M_{1}}\left({\bf \hat{g}}_{ij}^{}*{\bf \vec{h}}_{ij}^{\star 1}\right)^H\left({\bf \hat{g}}_{i'j}^{}*{\bf \vec{h}}_{i'j}^{\star 1}\right)\right|. \nonumber
\end{align}
in which,
\begin{align}\label{eq:Vector_h_ij}
{\bf \vec{h}}_{ij}^{\star 1}=\frac{\left({{\pmb \Phi}}_{ij}^{\star 1}\right)^{}\left\Vert {\bf \hat{h}}_{ij}^{1}\right\Vert }{\left(1-\sqrt{\psi}\right)},
\end{align}
where ${{\pmb \Phi}}_{ij}^{\star 1}$
can be obtained by computing the orthonormal eigenvector corresponding
to the largest eigenvalue of the matrix $\left({\bf \hat{G}}_{ij}^{}\left({\bf \hat{G}}_{ij}^{}\right)^{H}\dfrac{\left\Vert {\bf \hat{h}}_{ij}^{1}\right\Vert ^{2}}{\left(1-\sqrt{\psi}\right)^{2}}\right)$,
and 
\begin{align}\label{eq:Toe}
&{\bf \hat{G}}_{ij}^{} \in \mathbb C^{(2L-1)\times L}= \nonumber \\
&\left[ {\begin{array}{*{20}{c}}
   {{\hat g_{{{ij}}}}\left[ 1 \right]} & 0 & 0 & 0 & 0  \\
   {{\hat g_{{{ij}}}}\left[ 2 \right]} & {{\hat g_{{{ij}}}}\left[ 1 \right]} &  \cdots  &  \vdots  &  \vdots   \\
   {{\hat g_{{{ij}}}}\left[ 3 \right]} & {{\hat g_{{{ij}}}}\left[ 2 \right]} &  \cdots  & 0 & 0  \\
    \vdots  & {{\hat g_{{{ij}}}}\left[ 3 \right]} &  \cdots  & {{\hat g_{{{ij}}}}\left[ 1 \right]} & 0  \\
   {{\hat g_{{{ij}}}}\left[ {L - 1} \right]} &  \vdots  &  \cdots  & {{\hat g_{{{ij}}}}\left[ 2 \right]} & {{\hat g_{{{ij}}}}\left[ 1 \right]}  \\
   {{\hat g_{{{ij}}}}\left[ L \right]} & {{\hat g_{{{ij}}}}\left[ {L - 1} \right]} &  \vdots  &  \vdots  & {{\hat g_{{{ij}}}}\left[ 2 \right]}  \\
   0 & {{\hat g_{{{ij}}}}\left[ L \right]} &  \cdots  & {{\hat g_{{{ij}}}}\left[ {L - 2} \right]} &  \vdots   \\
   0 & 0 &  \cdots  & {{\hat g_{{{ij}}}}\left[ {L - 1} \right]} & {{\hat g_{{{ij}}}}\left[ {L - 2} \right]}  \\
    \vdots  &  \vdots  &  \vdots  & {{\hat g_{{{ij}}}}\left[ L \right]} & {{\hat g_{{{ij}}}}\left[ {L - 1} \right]}  \\
   0 & 0 & 0 &  \cdots  & {{\hat g_{{{ij}}}}\left[ L \right]}  \\
\end{array}} \right].
\end{align}
\end{lem}

\begin{IEEEproof}
Please refer to Appendix C.
\end{IEEEproof}

Accordingly, the proposed upper-bound on $P_{(isi)}{}_{j}^{1}$ can
be calculated as
\begin{align}\label{eq:Optimum2}
    \mathcal Pu_{(isi)}{}_{j}^{1}&= \sum\limits _{i=1}^{M_{1}}\left\Vert \sqrt{p_{j}^{1}}{\bf \hat{g}}_{ij}^{}*{\bf \vec{h}}_{ij}^{\star 1}\right\Vert ^{2} - \mathcal Pl_{(sig)}{}_{j}^{1}\\
     &+\left| \sum\limits _{i=1}^{M_{1}}\sum\limits _{{i=1\atop i'\ne i}}^{M_{1}}\left(\sqrt{p_{j}^{1}}{\bf \hat{g}}_{ij}^{}*{\bf \vec{h}}_{ij}^{\star 1}\right)^H\left(\sqrt{p_{j'}^{1}}{\bf \hat{g}}_{i'j}^{}*{\bf \vec{h}}_{i'j}^{\star 1}\right)\right|.  \nonumber
\end{align}

\subsection{Worst-case upper bound on the co-tier interference and objective function}
Similarly, the upper bound on the co-tier interference may be presented as follows
\begin{align}\label{eq:Optimum3}
\mathcal Pu_{(co)}{}_{j}^{1} & =\sum\limits _{{j'=1\atop j'\ne j}}^{N_{1}}\left(\sum\limits _{i=1}^{M_{1}}\left\Vert {\bf \hat{g}}_{ij'}^{}*{\bf \vec{h}}_{ij}^{\star 1}\right\Vert ^{2}p_{j'}^{1}\right) \nonumber \\
 & +\sum\limits _{{j'=1\atop j'\ne j}}^{N_{1}}\left| \sum\limits _{i=1}^{M_{1}}\sum\limits _{{i=1\atop i'\ne i}}^{M_{1}}\left({\bf \hat{g}}_{ij'}^{}*{\bf \vec{h}}_{ij}^{\star 1}\right)^{H}\left({\bf \hat{g}}_{i'j'}^{}*{\bf \vec{h}}_{i'j}^{\star 1}\right)p_{j'}^{1}\right|.
\end{align}
and the upper bound on the objective function is developed as
\begin{align} \label{eq:Optimum4}
&\max\limits _{{\bf e}_{in}^{10} \in \mathcal F_{in}^{10}}\sum\limits _{j=1}^{N_{1}}\left(\sum\limits _{n=1}^{N_{0}}\left\Vert \sum\limits _{i=1}^{M_{1}}{\bf \hat{g}}_{ij}^{}*{\bf h}_{in}^{10}\right\Vert ^{2}p_{j}^{1}\right)\nonumber
\\
 & \quad=\sum\limits _{j=1}^{N_{1}}\left(\sum\limits _{n=1}^{N_{0}}\sum\limits _{i=1}^{M_{1}}\left\Vert {\bf \hat{g}}_{ij}^{}*{\bf \vec{h}}_{in}^{\star 10}\right\Vert ^{2}p_{j}^{1}\right)\nonumber
 \\
 & \quad+\sum\limits _{j=1}^{N_{1}}\left|\sum\limits _{n=1}^{N_{0}}\sum\limits _{i=1}^{M_{1}}\sum\limits _{{i'=1\atop i'\ne i}}^{M_{1}}\left({\bf \hat{g}}_{ij}^{}*{\bf \vec{h}}_{in}^{\star 10}\right)^{H}\left({\bf \hat{g}}_{i'j}^{}*{\bf \vec{h}}_{i'n}^{\star 10}\right)p_{j}^{1}\right|\nonumber
 \\
& \quad =\Omega\left({\bf p}^{1}\right).
\end{align}
Following the worst-case approach and results from \eqref{eq:Optimum1} and \eqref{eq:Optimum2}, \eqref{eq:Optimum3}, and \eqref{eq:Optimum4}, the problem OP$_6$ can be approximately relaxed as problem OP$_6$ 
\begin{align}\label{eq:SimularProblem}
{\text{OP$_6$:~} \underset{\bf p^{1}}{\text{min}}} & \quad\Omega_{}\left({\bf p}^{1}\right)\nonumber
\\
\text{s.t.} & \quad\frac{\mathcal Pl_{(sig)}{}_{j}^{1}}{\mathcal Pu_{(isi)}{}_{j}^{1}+\mathcal Pu_{(co)}{}_{j}^{1}+P_{(tol)}{}_{j}^{01}+\left\Vert {\bf n}_{F}\right\Vert ^{2}}\ge\gamma_{j}^{k}.
\end{align}
This problem can be solved by the use of a similar approach adopted to tackle OP$_1$.

\section{Numerical Results}\label{sec:NumericalResult}
The impact of the proposed power allocation strategy on the system performance is analyzed in this section.
Without other statements, the system parameters are set following the ITU-R channel model \cite{Her2012} which is applicable for vehicular and indoor communication environments where the macrocell and femtocell are implemented, respectively. 

\begin{table}
\begin{centering}
\caption{{ITU Indoor Office \cite{Her2012}}}
\par\end{centering}
\centering{}%
\begin{tabular}{|l|c|c|}
\hline
\textbf{Tap} &
\textbf{Relative Delay (ns)}&
\textbf{Average Power (dBm)}
\tabularnewline
\hline
1 & 0 & 0
\tabularnewline
\hline
2 & 50 & -3
\tabularnewline
\hline
3 & 100 & -10
\tabularnewline
\hline
4 & 170 & -18
\tabularnewline
\hline
5 & 290 & -26
\tabularnewline
\hline
6 & 310 & -32
\tabularnewline
\hline
\end{tabular}
\end{table}

\begin{table}
\begin{centering}
\caption{{ITU Vehicular \cite{Her2012}}}
\par\end{centering}
\centering{}%
\begin{tabular}{|l|c|c|}
\hline
\textbf{Tap} &
\textbf{Relative Delay (ns)}&
\textbf{Average Power (dBm)}
\tabularnewline
\hline
1 & 0 & 0
\tabularnewline
\hline
2 & 310 & -1
\tabularnewline
\hline
3 & 710 & -9
\tabularnewline
\hline
4 & 1090 & -10
\tabularnewline
\hline
5 & 1730 & -15
\tabularnewline
\hline
6 & 2510 & -20
\tabularnewline
\hline
\end{tabular}
\end{table}

In the HetNet, the radii of MBS and FBS are $d_m = 300$ m and $d_f = 30$ m respectively. FBS is uniformly
distributed in a circle of $d_{mf} = 100$ m far from MBS. MUs and FUs are also
uniformly distributed in the served areas of MBS and FBS, respectively.
{
Specifically, the marcocell, femtocell and the the cross-tier channels are considered as the ITU vehicular (Table II), the ITU indoor (Table III), the ITU outdoor to indoor (Table IV) models, repectively.
}Moreover, we assume that
\begin{itemize}
\item The outdoor link pathloss exponent is set to 4, thus each tap of MBS-to-MU links is $\mathcal{CN}(0,\left|{\bf h}_{mn}^0[l]\right|^2)$ where $\left|{\bf h}_{mn}^0[l]\right|^2 = {\sigma _{0mn,l}^2}/{d_{0n}^4}$ and $d_{0n}$ is the distance between MBS and the $n$-th MU $(0 \le d_{0n} \le d_m)$.
\item The indoor link pathloss exponent is set to 3, thus each tap of FBS-to-FU links is $\mathcal{CN}(0,\left|{\bf h}_{ij}^1[l]\right|^2)$ where $\left|{\bf h}_{ij}^1[l]\right|^2 = {\sigma _{1ij,l}^2}/{d_{1j}^3}$ and $d_{1j}$ is the distance between FBS and the $j$-th FU $(0 \le d_{1j} \le d_f)$.
\item The outdoor-to-indoor link pathloss exponent is set to 3.5, thus each tap of MBS-to-FU links is $\mathcal{CN}(0,\left|{\bf h}_{mj}^{01}[l]\right|^2)$ where $\left|{\bf h}_{mj}^{01}[l]\right|^2 = {\sigma _{01mj,l}^2}/{d_{01j}^{3.5}}$ and $d_{01j}$ is the distance between MBS and the $j$-th FU $(0 \le d_{01j} \le d_f)$.
\item The indoor-to-outdoor links are assumed to be similar to the outdoor-to-indoor links.
\end{itemize}
{
It is worth noting that there exist no correlations between the channels and between their taps.
}

\begin{table}
\begin{centering}
\caption{{ITU Outdoor to Indoor and Pedestrian \cite{Her2012}}}
\par\end{centering}
\centering{}%
\begin{tabular}{|l|c|c|}
\hline
\textbf{Tap} &
\textbf{Relative Delay (ns)}&
\textbf{Average Power (dBm)}
\tabularnewline
\hline
1 & 0 & 0
\tabularnewline
\hline
2 & 110 & -9.7
\tabularnewline
\hline
3 &190 & -19.2
\tabularnewline
\hline
4 & 410 & -22.8
\tabularnewline
\hline
\end{tabular}
\end{table}

\begin{table}
\begin{centering}
\caption{IMPORTANT PARAMETERS}
\par\end{centering}
\centering{}%
\begin{tabular}{|l|c|}
\hline
\textbf{Parameters} &
\textbf{System values}
\tabularnewline
\hline
Number of taps, $L$ &
6\tabularnewline
\hline
Number of antennas at \gls{mbs}, $M_0$&
4\tabularnewline
\hline
Number of antennas at \gls{fbs}, $M_1$ &
4\tabularnewline
\hline
Number of users at \gls{mbs}, $N_0$ &
2\tabularnewline
\hline
Number of users at \gls{fbs}, $N_1$ &
2-4\tabularnewline
\hline
Tolerable level of cross-interference, $P_{(tol)}{}_{j}^{01}$ &
-10 dBm\tabularnewline
\hline
Bandwidth, $W$&
20 MB\tabularnewline
\hline
\end{tabular}
\end{table}

For convenience, we set the SINR thresholds for FBS as $\left\{ \gamma_{j}^{1}\right\} _{j=1}^{N_{1}}=\gamma_{F}$
and for MBS as $\left\{ \gamma_{n}^{0}\right\} _{n=1}^{N_{0}}=\gamma_{M}$. In addition, the Gaussian noise power is set to $10^{-12}$ W
and other parameters are adjusted as listed in Table VI.

\subsection{The Proposed Power Allocation Strategy}
\begin{figure}
\includegraphics[scale=0.55]{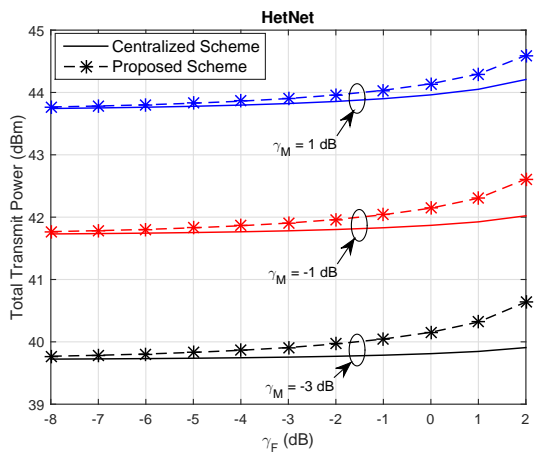}
\caption{Performance of power allocation schemes.}
\label{fig:5}
\end{figure}
In Fig.~\ref{fig:5}, we investigate the transmit power of the HetNet system
with the proposed optimization approach, and the scheme using centralized manner.
The simulation is carried out with $1000$
random locations of MBS, FBS and users in the considered HetNet environment.
Since MUs are frequently located far from its own base station, the distance between MBS and MUs is much larger compared with the distance between FBS and FUs. Therefore, the amount of transmit power allocated for MUs constitutes a major part of the total transmit power of HetNet. As a result, the total transmit power slightly increases when the SINR threshold of FUs scales up.  Fig.~\ref{fig:5} also shows the comparison between the centralized and the proposed approaches.
In more details,
the power gap at $\gamma_{F}=2$ dB is roughly $0.6$ dB,
$0.5$ dB and $0.4$ dB in cases of $\gamma_{M}=1$ dB, $-1$ dB and $-3$ dB, respectively.
Furthermore, the gap scales up as the SINR threshold of FUs increases. This is because the FBS aims at minimizing the interference to MUs, instead of solely minimizing the transmit power as in the case of centralized approach. More specifically, the proposed scheme sacrifices an additional amount of transmit power for (i) a much smaller amount of required signaling information and (ii) a reduced interference to MU. Note that only $\left\{P^\star_{(cross)}{}_{n}^{0}\right\}_{n=1}^{N_0}$ is sent via backhaul to the
macrocell instead of ${\bf p}^{\star1}$, $\left\{ {\bf h}_{ij}^1\right\} _{i=1,\: j=1}^{M_{1},\: N_{1}}$
and $\left\{ {\bf h}_{in}^{10}\right\} _{j=1,\: n=1}^{N_{1},\: N_{0}}$, the signaling overhead is significantly reduced. Indeed, based on Fig.~\ref{fig:5}, one can conclude that the proposed optimization algorithm can achieve tight results to the centralized strategy which validates and verifies our strategy.

\begin{figure}
\includegraphics[scale=0.55]{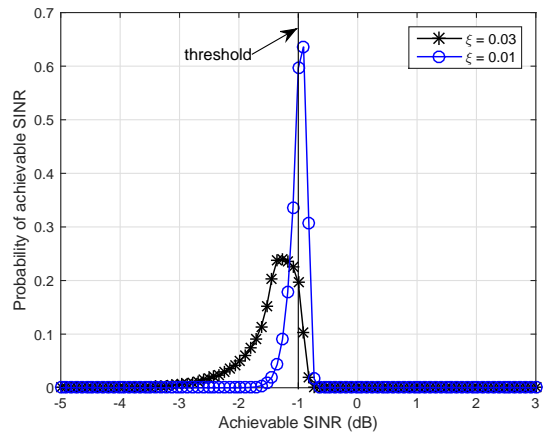}
\caption{{Impact of imperfect $\left\{ {\bf h}_{in}^{10}\right\} _{j=1,\: n=1}^{N_{1},\: N_{0}}$ on the performance at a MU.}}
\label{fig:10}
\end{figure}

{Moreover, in practice, obtaining the perfect information of cross channels, such as $\left\{ {\bf h}_{in}^{10}\right\} _{i=1,\: n=1}^{M_{1},\: N_{0}}$ and $\left\{ {\bf h}_{mj}^{01}\right\} _{m=1,\: j=1}^{M_{0},\: N_{1}}$, is a challenge. Although many previous works  \cite{Joe2011, Duy2015,Cha2009} assume that the information is available at the base station or the computation node,
this motivates us to further consider the impact of imperfect $\left\{ {\bf h}_{in}^{10}\right\} _{j=1,\: n=1}^{N_{1},\: N_{0}}$ on the SINR performance achieved at MUs. The latter represents the case in which FBS imperfectly estimate the CSI of $\left\{ {\bf h}_{in}^{10}\right\} _{j=1,\: n=1}^{N_{1},\: N_{0}}$, and then it sends the "inaccurate" information of minimized interference to MBS.
In this concern, we use a general imperfect channel model as $\hat {\bf h}_{in}^{10} = {\bf h}_{in}^{10} + {\bf e}_{in}^{10}$, where ${\bf e}_{in}^{10}$ is a channel uncertainty and $\left\|{\bf e}_{in}^{10}\right\|^2 \le \xi{\bf h}_{in}^{10}$. In fact, the inexact information of $\left\{ {\bf h}_{in}^{10}\right\} _{j=1,\: n=1}^{N_{1},\: N_{0}}$ might lead to the fact that the achieved SINR performance at each MU is not guaranteed to meet the preset threshold. In Fig.~\ref{fig:10}, the impact of imperfect CSI is shown in terms of the probability of the achievable SINR at a MU. As expected, given the threshold of $-1$ dB, it is clear that the outage probability increases when the error component scales up. 
}

\begin{figure}
\emph{\includegraphics[scale=0.55]{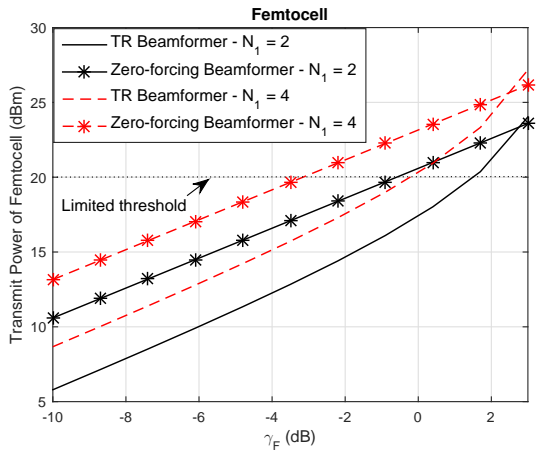}}
\caption{A comparison between TR and zero-forcing.}
\label{fig:6}
\end{figure}

\subsection{Comparison between TR and zero-forcing techniques}
In Fig.~\ref{fig:6}, we compare the effectiveness of TR-based beamformer with
that of zero-forcing-based beamformer (\textit{Algorithm} 1).
Concerning channels, the ITU-R indoor model is still utilized, however, we arrange a distance of 15 meters between FBS and FUs.
From Fig.~\ref{fig:6}, it is visible that for a transmit power range lower than $23$ dBm and $25$ dBm for cases of $N_1=2$ users and $N_1=4$ users, respectively, TR beamforming outperforms
zero-forcing one and converse holds for the entire transmit power regions. 
Specifically, it can be explained that
zero-forcing scheme mainly deals with canceling the ISI, co-tier interference and cross-tier interference, whereas TR technique
aims at both focusing signal power on the central tap, and reducing the ISI and co-tier interferences. As a result, the interference is ineliminable completely in TR-applying systems, and the interference power increases as transmit power scales up. Therefore, there exist working ranges in which either TR or zero-forcing techniques can dominate the other one regarding SINR metric.
In practice, however, FBS is a low-power cellular station whose transmit power is limited to $20$dBm in order to curb the effects of co- and cross-tier interferences \cite{Zah2013}.
Indeed, in perspective of a small-cell system configured with a limited level of transmit power, such as femtocell operating environments, it can be concluded that the TR technique is more desirable than the zero-forcing.

\subsection{Worst-case Optimization Problem and Performance of Proposed Upper-bounds}
\begin{figure}
\includegraphics[scale=0.55]{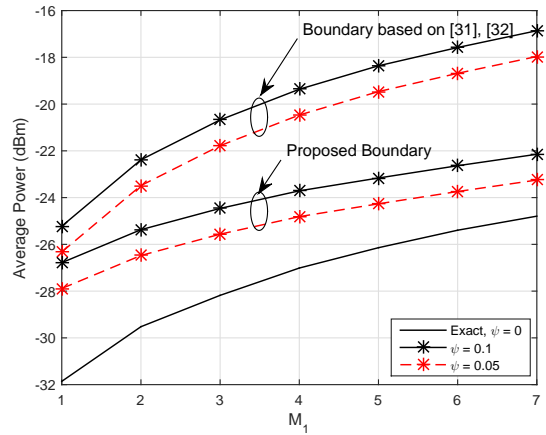}
\caption{A comparison between worst-case upper-bounds.}
\label{fig:7}
\end{figure}
In Section \ref{sec:WorstCase}, we have proposed a novel worst-case upper-bound to provide
a greater solution to the robust downlink power allocation for the TR femtocell. First, for simplicity, we observe the term $\left\Vert \sum\limits _{i=1}^{M_{1}}{\bf \hat{g}}_{ij}*{\bf h}_{ij}^{1}\right\Vert ^{2}$ to evaluate the performance of the proposed boundary and Young\textquoteright{}s inequality-based
boundary.
Simulations are carried out in indoor channels with a fixed pathloss similar to the prior investigation. Fig.~\ref{fig:7} clearly demonstrates that the proposed boundary is approximately $5$dB tighter than
Young\textquoteright{}s inequality-based
boundary at $M_{1}=4$ for both cases of error factor $\psi = 0. 05, 0.1$.

\begin{figure}
\emph{\includegraphics[scale=0.55]{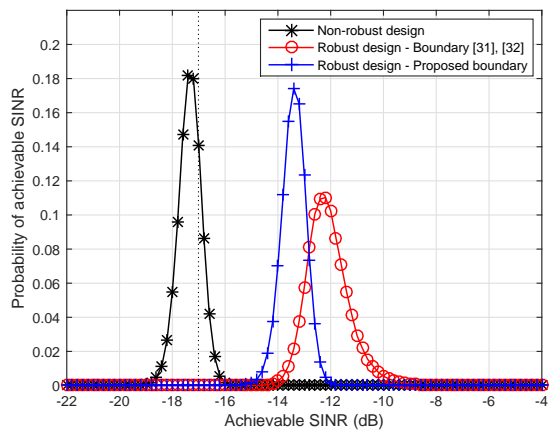}}
\caption{Probability distribution per a femtocell user.}
\label{fig:8}
\end{figure}

\begin{figure}
\emph{\includegraphics[scale=0.55]{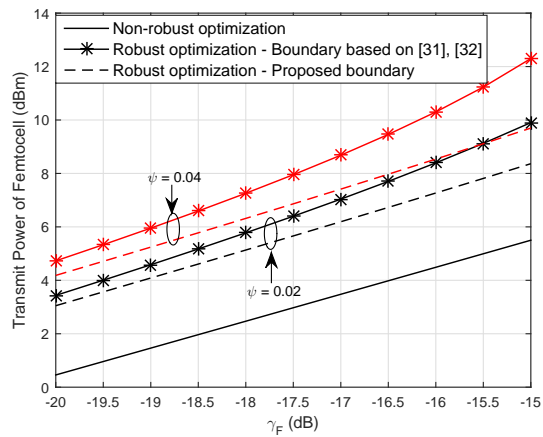}}
\caption{Transmit power of femtocell with worst-case upper-bounds.}
\label{fig:9}
\end{figure}

Concerning the worst case robust design, Fig.~\ref{fig:8} and Fig.~\ref{fig:9} demonstrate the transmission power in terms of worst-case approaches using:
(i) the Young\textquoteright{}s inequality-based boundary as well
as (ii) the proposed boundary, under similar error conditions. In more details, Fig.~\ref{fig:8}
exhibits the probability distribution of achievable SINR per FU
obtained by non-robust design (OP$_1$) and robust design (OP$_6$). This result is achieved by use of $100000$
randomly generated realizations for the simulation. For the channel error factor $\psi=0.04$, our simulation shows
that two robust designs can keep the outage probability equal to $0$ by accounting the
worst possible error case. Nevertheless, the probability distribution
of the design employing the proposed boundary is closer to the preset
threshold than that of Young's inequality-based design. It can be explained that FBS
can save the radiated energy in obtaining the same desired SINR performance
when applying the proposed boundary.

Moreover, as a reference, we plot the average transmit power of non-robust and robust manners in Fig.~\ref{fig:9} and exhibit the amount of additional power that is needed to achieve a zero outage probability. It can be concluded that the advantage of our boundary makes the transmit power allocation more effective in the \emph{worst-case approach} by curbing the waste of power transmission.

\section{Conclusion}\label{sec:Conclusion}
In this paper, we propose the application of \gls{tr} technique to femtocell networks, and a novel power allocation scheme for the considered HetNet in which the backhaul connection provides only a limited throughput for signaling exchange.
In perfect channel estimation cases, we tackle the beamformer designs and optimization
problems of downlink power control for both macrocell and femtocell
over frequency selective fading channels. 
Our analysis shows that the proposed allocation schemes require a higher increment in transmit power compared to the conventional approach but demands a lower amount of signaling exchange between the MBS and FBS. This important advantage makes our approach very promising to deal with limited backhaul connection drawbacks.
Furthermore, under imperfect \gls{csi} assumption, we tackle the robust design following the worst-case approach. To relax the original formulation into a solvable convex problem, the worst-case boundaries of concerning components are derived. In particular, we propose the novel tighter worst-case upper-bound of the ISI, co-tier interference and objective function to improve the system performance. Moreover, numerical results demonstrate that the TR technique outperforms the zero-forcing beamforming over femtocell working environments.

\appendices
\section{Proof of Lemma 1}
To solve this problem OP$_{1}$, we start with introducing a new parameter for the objective function as follows
\begin{align}
\sum\limits _{j=1}^{N_{1}}\left(\sum\limits _{n=1}^{N_{0}}\left\Vert \sum\limits _{i=1}^{M_{1}}{\bf g}_{ij}*{\bf h}_{in}^{10}\right\Vert ^{2}p_{j}^{1}\right)
= {\hat{\pmb \eta}}^T {\bf p}^{1},
\end{align}
where ${\hat{\pmb \eta}}$ are defined in \emph{Lemma} 1. 

Next, the expression of $\text{SINR}_{j}^{1}$ in \eqref{eq:SINR_j1} can be re-written as 
\begin{align}
&\text{SINR}_{j}^{1}\left(L,{\bf \; p}^{1},\;\left\{ {\bf g}_{ij}^{1}\right\} _{i=1,j=1}^{M_{1},N_{1}},{P_{(cross)}{}_{j}^{1}}\right) \\ \nonumber
&=\frac{ \left(\left|\left(\sum\limits _{i=1}^{M_{1}}{\bf g}_{ij}*{\bf h}_{ij}^1\right)\left[L\right]\right|^{2}-\gamma_{j}^{1}\sum\limits _{{l=1\atop l\ne L}}^{2L-1}\left|\left(\sum\limits _{i=1}^{M_{1}}{\bf g}_{ij}*{\bf h}_{ij}^{1}\right)\left[l\right]\right|^{2}\right)p_j^1   } {{P_{(co)}{}_{j}^{1}}+{P_{(cross)}{}_{j}^{1}}+\left\Vert {\bf n}_{F}[\beta]\right\Vert ^{2}}.
\end{align}

Accordingly, in form of matrix-vector notation, the OP$_1$ can be formulated as follows
\begin{align}\label{eq:SUB-OP11}
    \begin{aligned}
        {\text{OP$_{1-1}$:~} \underset{\bf p^{1}}\min}\quad & {\hat{\pmb \eta}}^T{ {\bf p}}^{1}
        \\
        \text{s.t.}\quad & { {\bf p}}^{1} \succeq {\bf DB}{ {\bf p}}^{1}  + {\bf z}, 
    \end{aligned}
\end{align}
in which, the structures of ${\bf D}$, ${\bf B}$, and ${\bf z}$ are defined in \emph{Lemma} 1.

Considering the objective function of OP$_{1-1}$, it is a fact that
\begin{align}
{\hat{\pmb \eta}}^T {\bf p}^{1} = \left\| {\hat{\pmb \eta}} \right\| {\rm Tr}\left({\rm diag}({\pmb \eta}){ {\bf p}}^{1} \right),
\end{align}
where ${\pmb \eta}$ is the normalized form of ${\hat{\pmb \eta}}$ which can be computed as ${\pmb \eta} = \dfrac{\hat{\pmb \eta}}{\left\| {\hat{\pmb \eta}} \right\|}$.

On this basis, we aim at deriving the constraint of OP$_{1-1}$ as a function of $\left({\rm diag}({\pmb \eta}){ {\bf p}}^{1} \right)$. After some manipulations, the constraint can be shown in a new equivalent formulation as
\begin{align}
&{ {\bf p}}^{1} \succeq {\bf DB}{ {\bf p}}^{1}  + {\bf z} \nonumber \\
&\Leftrightarrow {\rm diag}({\pmb \eta}){ {\bf p}}^{1} \succeq {\rm diag}({\pmb \eta} ){\bf D} \left({\bf B}\circ ({\pmb \eta}^{-1} {\bf 1}_{N_{1}\times {1}}^T )\right) {\rm diag}({\pmb \eta}){ {\bf p}}^{1} + {\bf z}.
\end{align}

Hence, the OP$_{1-1}$ can be re-formulated as
\begin{align}\label{eq:SUB-OP12}
    \begin{aligned}
        {\text{OP$_{1-2}$:~} \underset{\bf p^{1}}\min}\quad & {\left\| {\hat{\pmb \eta}} \right\|}{\rm Tr}\left({\rm diag}({\pmb \eta}){ {\bf p}}^{1} \right)
        \\
        \text{s.t.}\quad & {\rm diag}({\pmb \eta}){ {\bf p}}^{1} \succeq {\rm diag}({\pmb \eta} ){\bf D}\left({\bf B}\circ ({\pmb \eta}^{-1} {\bf 1}_{N_{1}\times {1}}^T )\right)\\
& \quad\quad\quad\quad \quad\quad \quad\quad\quad \quad\times {\rm diag}({\pmb \eta}){ {\bf p}}^{1} + {\bf z}. 
    \end{aligned}
\end{align}

Therefore, the closed-form expression of ${ {\bf p}}^{\star 1}$ can be given by
\begin{align}
{ {\bf p}}^{\star 1}={\rm diag}({\pmb \eta})^{-1}\left[{\bf I}_{N_{1}}-{\rm diag}({\pmb \eta} ){\bf D}\left({\bf B}\circ ({\pmb \eta}^{-1} {\bf 1}_{N_{1}\times {1}}^T )\right)  \right]^{-1}{\bf z}.
\end{align}

Relying on Perron-Frobenius theory, the optimal value ${ {\bf p}}^{\star 1}$ is guaranteed to be a nonnegative vector if and only if the spectral radius of ${\rm diag}({\pmb \eta} ){\bf D}\left({\bf B}\circ ({\pmb \eta}^{-1} {\bf 1}_{N_{1}\times {1}}^T )\right)$ is less than unity.

Furthermore, in a special case where $\eta_1 = \eta_2 = ... = \eta_{N_1} = \eta$, the above problem can be written in a simplified form as 
\begin{align}\label{eq:SUB-OP13}
    \begin{aligned}
        {\text{OP$_{1-3}$:~} \underset{\bf p^{1}}\min}\quad & \eta{\bf I}_{N_1}^T {\bf p}^{1}
        \\
        \text{s.t.}\quad & { {\bf p}}^{1} \succeq {\bf DB}{ {\bf p}}^{1}  + {\bf z}. 
    \end{aligned}
\end{align}
Then, the optimal solution of OP$_{1-2}$ can be given as
\begin{equation}\label{eq:SOL-OP13}
{{\bf p}^{\star 1}}=\left({\bf I}-{\bf DB}\right)^{-1}{\bf z}.
\end{equation} 
One can see that the result of \eqref{eq:SOL-OP13} is in agreement with that of the work in \cite{Cha2009}. In final, the proof is completed.

\section{Proof of Lemma 2}
According to \eqref{eq:fol1}, we can write down that
\begin{align}
& \underset{{\bf e}_{ij}^{1} \in \mathcal F_{ij}^1}{\min} \left|\sum\limits _{i=1}^{M_{1}}({\hat{\bf g}}_{ij}^{}*{ {\bf h}}_{ij}^{1})[L]\right|^{2} \nonumber \\
&=\underset{{\bf e}_{ij}^{1}\in \mathcal F_{ij}^1}{\min} \frac{\left| \sum\limits _{i=1}^{M_{1}} \left( \left\Vert {\bf h}_{ij}^{1}\right\Vert ^{2}+({\bf e}_{ij}^{1})^{H}{\bf h}_{ij}^{1} \right)\right|^{2}}{\sum\limits _{i=1}^{M_{1}}\left\Vert {\bf \hat{h}}_{ij}^{1}\right\Vert ^{2}} \nonumber\\
&\mathop  = \limits^{(a)} \frac{ \left| \sum\limits _{i=1}^{M_{1}} \left( \left\Vert {\bf e}_{ij}^{1}\right\Vert ^{2}  \left( \dfrac{1-\sqrt{\psi}}{\psi} \right)  \right) \right|^{2}} {\sum\limits _{i=1}^{M_{1}}\left\Vert {\bf \hat{h}}_{ij}^{1}\right\Vert ^{2}}.
\end{align}
The equality in $(a)$ is due to the fact that the quantity of $\left( \left\Vert {\bf h}_{ij}^{1}\right\Vert ^{2}+({\bf e}_{ij}^{1})^{H}{\bf h}_{ij}^{1}\right)$ reaches the minimum if and only if ${\bf e}_{ij}^{1} = - \sqrt{\psi}{\bf h}_{ij}^{1}$. 

Next, by substituting ${\bf e}_{ij}^{1} = - \sqrt{\psi}{\bf h}_{ij}^{1}$ to \eqref{eq:fol2}, we can obtain the following derivation
\begin{align}\label{eq:fol3}
\left\Vert {\bf e}_{ij}^{1}\right\Vert ^{2} = \dfrac{\psi}{(1-\sqrt{\psi})^2}\left\Vert {\bf \hat{h}}_{ij}^{1}\right\Vert ^{2}.
\end{align}
Thus, we infer that
\begin{align}
&\underset{{\bf e}_{ij}^{1}\in \mathcal F_{ij}^1}{\min} \left|\sum\limits _{i=1}^{M_{1}}({\hat{\bf g}}_{ij}^{}*{ {\bf h}}_{ij}^{1})[L]\right|^{2} =\dfrac{\sum\limits _{i=1}^{M_{1}}\left\Vert {\bf \hat{h}}_{ij}^{1}\right\Vert ^{2}}{(1-\psi)^2}.
\end{align}

On the other hand, the estimated signal power can be given as
\begin{align}
\left|\sum\limits _{i=1}^{M_{1}}({\hat{\bf g}}_{ij}^{}*{ \hat{\bf h}}_{ij}^{1})[L]\right|^{2}= {\sum\limits _{i=1}^{M_{1}}\left\Vert {\bf \hat{h}}_{ij}^{1}\right\Vert ^{2}}.
\end{align}

Accordingly, the worst-case lower-bound of signal
power component can be constituted as
\begin{align}\label{eq:Optimum1}
{\mathcal Pl}_{(sig)}{}_{j}^{1}=\left|\sum\limits _{i=1}^{M_{1}}({\hat{\bf g}}_{ij}^{}*{ \hat{\bf h}}_{ij}^{1})[L]\right|^{2} \frac{{p_{j}^{1}}}{\left(1-\sqrt{\psi}\right)^{2}}.
\end{align}
This completes the proof of \emph{Lemma} 2.

\section{Proof of Lemma 3}
We consider the worst-case upper-bound of
$\left\Vert {\bf \hat{g}}_{ij}^{}*{\bf h}_{ij}^{1}\right\Vert ^{2}$
for simplicity. To investigate a tighter upper-bound
than the value suggested in \cite{You1912,Bec1975},
we introduce a novel approach in which we start with letting ${\bf \vec{h}}_{ij}^{1}$
be a virtual channel, and we aim at investigating the optimal value ${\bf \vec{h}}_{ij}^{\star 1}$
that makes $\left\Vert {\bf \hat{g}}_{ij}^{}*{\bf \vec{h}}_{ij}^{1}\right\Vert ^{2}$
achieve maximal quantity. Accordingly, the problem becomes
\begin{align}\label{eq:ConvoOptimization}
\underset{{\bf \vec{h}}_{ij}^{1}}{\max} & \quad\left\Vert {\bf \hat{g}}_{ij}^{}*{\bf \vec{h}}_{ij}^{1}\right\Vert ^{2} \nonumber \\
\text{s.t.} & \quad\left\Vert {\bf \vec{h}}_{ij}^{1}\right\Vert ^{2} \le \underset{{\bf e}_{ij}^{1}\in \mathcal F_{ij}^1}{\max} \left\Vert {\bf h}_{ij}^{1}\right\Vert ^{2}.
\end{align}

As observed, the problem \eqref{eq:ConvoOptimization} is hard to solve directly due to the convolution
operator. Thus, we factorize \eqref{eq:ConvoOptimization} into two steps. In the first
step, we let ${\bf \hat{G}}_{ij}^{}$ to be a $(2L-1)\times L$ Toeplitz
matrix form of ${\bf \hat{g}}_{ij}^{}$, i.e. \eqref{eq:Toe}

Additionally, we let ${\bf \Phi}_{ij}^{1} = \dfrac{{\bf \vec{h}}_{ij}^{1}}{\sqrt{\underset{{\bf e}_{ij}^{1}\in \mathcal F_{ij}^1}{\max} \left\Vert {\bf h}_{ij}^{1}\right\Vert ^{2}}}$.
Based on the representation
of convolution with Toeplitz matrix form, the above problem can be equivalently re-formulated as
\begin{align}\label{eq:ConvoOptimization2}
\underset{{\bf \Phi}_{ij}^{1}}{\max} & \quad\left\Vert \left( {\sqrt{\underset{{\bf e}_{ij}^{1}\in \mathcal F_{ij}^1}{\max} \left\Vert {\bf h}_{ij}^{1}\right\Vert ^{2}}}{\bf \hat{G}}_{ij} \right){\bf \Phi}_{ij}^{1}\right\Vert ^{2} \nonumber \\
\text{s.t.} & \quad { \left\Vert {\bf \Phi}_{ij}^{1}\right\Vert ^{2}} \le 1.
\end{align}

The closed-form derivation of optimal solution ${{\pmb \Phi}}_{ij}^{\star 1}$
can be obtained by computing the orthonormal eigenvector corresponding
to the largest eigenvalue of the matrix $\left({\bf \hat{G}}_{ij}^{}\left({\bf \hat{G}}_{ij}^{}\right)^{H}{\underset{{\bf e}_{ij}^{1}\in \mathcal F_{ij}^1}{\max} \left\Vert {\bf h}_{ij}^{1}\right\Vert ^{2}}\right)$.
Since the norm of ${{\pmb \Phi}}_{ij}^{\star 1}$ is equal to 1, to make
a fair normalization, we devise that
\begin{align}\label{eq:Vector_h_ij}
{\bf \vec{h}}_{ij}^{\star 1}=\frac{\left({{\pmb \Phi}}_{ij}^{\star 1}\right)^{}\left\Vert {\bf \hat{h}}_{ij}^{1}\right\Vert }{\left(1-\sqrt{\psi}\right)},
\end{align}
where ${\bf \vec{h}}_{ij}^{\star 1}$ is the optimal value regarding to the
achievement the maximum of $\left\Vert {\bf \hat{g}}_{ij}^{}*{\bf \vec h}_{ij}^{1}\right\Vert ^{2}$, and $\underset{{\bf e}_{ij}^{1}\in \mathcal F_{ij}^1}{\max} \left\Vert {\bf h}_{ij}^{1}\right\Vert ^{2} = \dfrac{\left\Vert {\bf \hat{h}}_{ij}^{1}\right\Vert ^{2}}{\left(1-\sqrt{\psi}\right)^{2}}$.

Therefore, we can obtain a new boundary from \eqref{eq:Vector_h_ij} as
\begin{align}
   &\underset{{\bf e}_{ij}^{1}\in \mathcal F_{ij}^1}{\max} \left\Vert \sum\limits _{i=1}^{M_{1}}{\bf \hat{g}}_{ij}^{}*{\bf h}_{ij}^{1}\right\Vert ^{2} \\
    & = \sum\limits _{i=1}^{M_{1}}\left\Vert {\bf \hat{g}}_{ij}^{}*{\bf \vec{h}}_{ij}^{\star 1}\right\Vert ^{2}+\left| \sum\limits _{i=1}^{M_{1}}\sum\limits _{{i=1\atop i'\ne i}}^{M_{1}}\left({\bf \hat{g}}_{ij}^{}*{\bf \vec{h}}_{ij}^{\star 1}\right)^H\left({\bf \hat{g}}_{i'j}^{}*{\bf \vec{h}}_{i'j}^{\star 1}\right)\right|. \nonumber
\end{align}
Thus, the proof of \emph{Lemma} 3 is completed.

\bibliographystyle{IEEEtran}
\bibliography{IEEEabrv,REF}

\end{document}